\documentclass{sig-alternate}
\usepackage{mathptmx} 

\usepackage{fancyhdr}
\usepackage[normalem]{ulem}
\usepackage[hyphens]{url}
\usepackage[sort,nocompress]{cite}
\usepackage[final]{microtype}
\usepackage[keeplastbox]{flushend}
\usepackage[bookmarks=true,breaklinks=true,letterpaper=true,colorlinks,citecolor=blue,linkcolor=blue,urlcolor=blue]{hyperref}

\usepackage{xcolor}
\usepackage{fancyhdr}
\usepackage[normalem]{ulem}

\usepackage{url}

\usepackage{breakurl}

\usepackage{booktabs}
\usepackage{svg}
\usepackage{subfig}
\usepackage{multirow}
\usepackage{multicol}
\usepackage{booktabs}
\usepackage{enumitem}
\usepackage{amsmath,amssymb, cite}
\usepackage{siunitx}
\usepackage{soul}
\usepackage{pifont}
\usepackage{newunicodechar}
\usepackage[skip=0pt]{caption}
\usepackage{mathtools}
\usepackage{adjustbox}

\newcommand{\E}[1]{{\mathbb{E}}\left[#1\right]}
\newunicodechar{✓}{\ding{51}}
\newunicodechar{✗}{\ding{55}}

\newcommand{\microsubmissionnumber}{XXX}
\fancypagestyle{firstpage}{
  \fancyhf{}
  
  \fancyhead[C]{\vspace{10pt}\normalsize{MICRO 2023 Submission
      \textbf{\#\microsubmissionnumber} -- Confidential Draft -- Do NOT Distribute!!}\\\vspace{-25pt}} 
  \fancyfoot[C]{\thepage}
}

\newcommand{\hadprod}[2]{{\vec{#1}}\circ{\vec{#2}}}
\newcommand{\hadprodnv}[2]{{#1}\circ{#2}}
\newcommand{\crecv}{\vec{c}_{recv}}
\newcommand{\cobt}{\vec{c}_{obt}}
\newcommand{\uc}{\vec{u}_{cost}}
\newcommand{\quant}{\vec{q}}
\newcommand{\zs}{\vec{Z}^s}
\newcommand{\zd}{\vec{Z}^d}
\newcommand{\cobtnv}{c_{obt}}

\newcommand{\yield}{\vec{y}}
\newcommand{\ucost}{\vec{u}_{cost}}

\newcommand{\cused}{\vec{c}_{used}}
\newcommand{\cusednv}{c_{used}}

\newcommand{\mappingquant}{\vec{U}}
\newcommand{\cbuilt}{\vec{c}_{built}}
\newcommand{\cdemand}{\vec{c}_{demand}}
\newcommand{\csold}{\vec{c}_{sold}}
\newcommand{\uben}{\vec{u}_{ben}}
\newcommand{\usc}{\vec{u}_{sc}}
\newcommand{\nre}{\vec{n}}
\newcommand{\ord}{\vec{o}}
\newcommand{\based}{\vec{b}}

\title{Understanding Interactions Between Chip Architecture and Uncertainties in Semiconductor Supply and Demand} 
\author{
Ramakrishna Kanungo \\
\email{kanungo3@illinois.edu} 
\and
Swamynathan Siva \\
\email{siva3@illinois.edu}
\and
Nathaniel Bleier \\
\email{nbleier3@illinois.edu}
\and
Muhammad Husnain Mubarik \\
\email{mubarik3@illinois.edu} 
\and 
Lav Varshney \\
\email{varshney@illinois.edu}
\and
Rakesh Kumar \\
\email{rakeshk@illinois.edu} 
}

\begin{document}
\maketitle
\pagestyle{plain}

\begin{abstract}
Mitigating losses from supply and demand volatility in the semiconductor supply chain and market has traditionally been cast as a logistics and forecasting problem. We investigate {\em how the architecture of a family of chips influences how it is affected by supply and demand uncertainties}. We observe that semiconductor supply chains become fragile, in part, due to single demand paths, where one chip can satisfy only one demand. Chip architects can enable multiple paths to satisfy a chip demand, which improves supply chain resilience. Based on this observation, we study composition and adaptation as architectural strategies to improve resilience to volatility and also introduce a third strategy of dispersion. These strategies allow multiple paths to satisfy a given chip demand.  We develop a model to analyze the impact of these architectural techniques on supply chain costs under different regimes of uncertainties and evaluate what happens when they are combined. We present several interesting and even counter-intuitive observations about the product configurations and market conditions where these interventions are impactful and where they are not. In all, we show that product redesign supported by architectural changes can mitigate nearly half of the losses caused by supply and demand volatility. As far as we know, this is the first such investigation concerning chip architecture. 
\end{abstract}

\section{Introduction}
Chip shortages are front and center these days. Mismatches between supply and demand continue to wreak havoc on our lives---affecting the supply of computers~\cite{comp} and automobiles~\cite{auto}, and are even implicated in  inflation~\cite{chips_cause_infl}.

\begin{figure*}[ht]
    \centering
    \captionsetup[subfloat]{captionskip=-1pt,nearskip=-1pt}
     \subfloat[\centering Fluctuations in the PC industry]{
        {\includegraphics[width=0.32\linewidth]{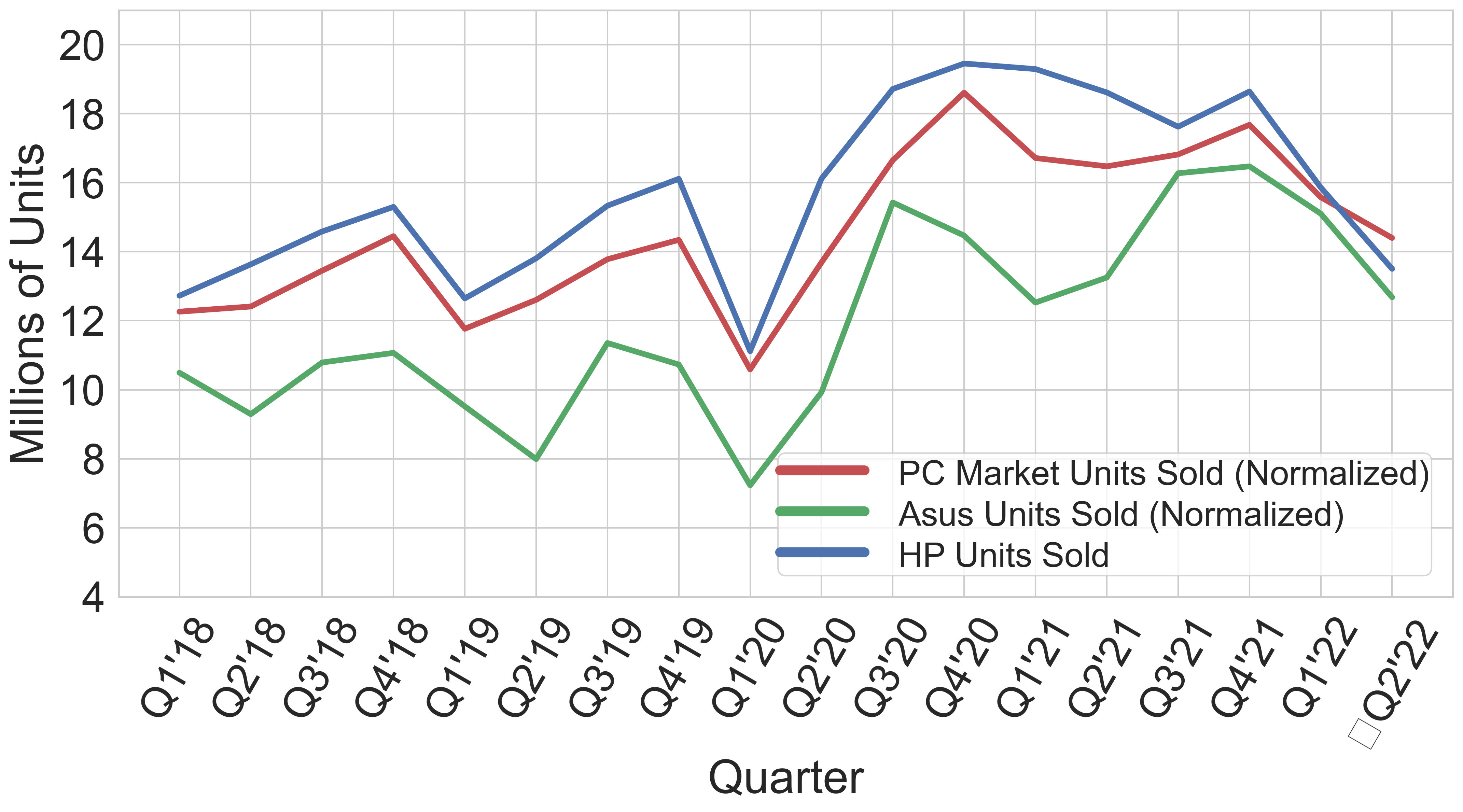}}
        \label{fig:PC_mar_fluct}
    } 
    \subfloat[\centering Semiconductor Revenues vs IDC and Gartner Projections]{
        {\includegraphics[width=0.32\linewidth]{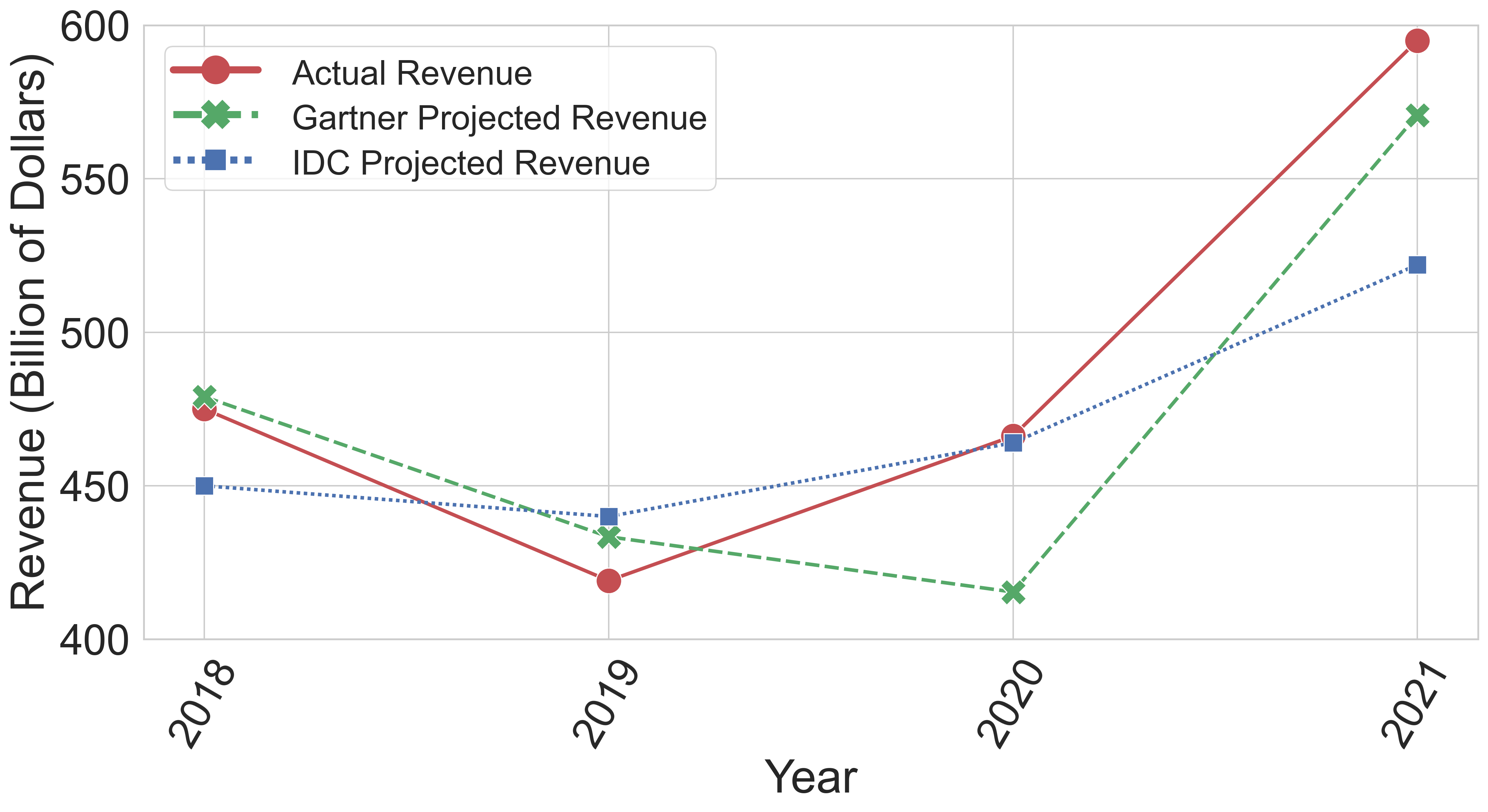}}
        \label{fig:Unc_in_forcast} 
    }
    \subfloat[\centering Semiconductor production Utilization and Inflation-adjusted Revenues]{
        {\includegraphics[width=0.32\linewidth]{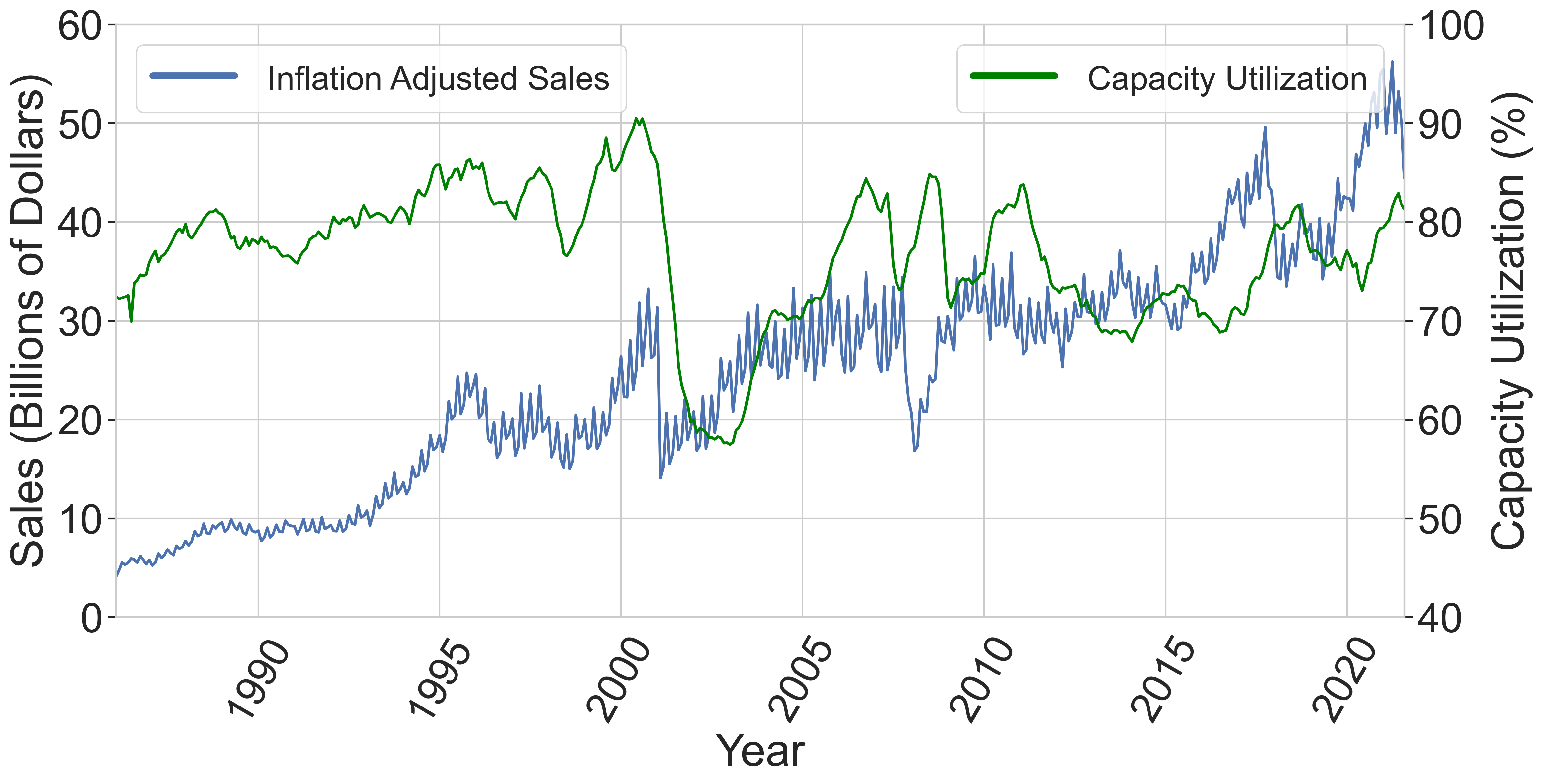}}
        \label{fig:Util_revenue} 
    }
    \label{fig:motivation}
    \caption{Semiconductor industry trends}
\end{figure*}

While public awareness of supply and demand swings affecting the semiconductor industry is at an all-time high, the problem is not new. 
Figure~\ref{fig:PC_mar_fluct} shows the number of PCs shipped per quarter by Hewlett-Packard, ASUS, and the market as a whole
~\cite{PC_shipments_2Q22, PC_shipments_1Q22, PC_shipments_4Q21, PC_shipments_3Q21, PC_shipments_2Q21, PC_shipments_1Q21, PC_shipments_4Q20, PC_shipments_3Q20, PC_shipments_2Q20, PC_shipments_1Q20, PC_shipments_4Q19, PC_shipments_3Q19, PC_shipments_2Q19, PC_shipments_1Q19}. 
We take the average of HP shipments and scale up/scale down the ASUS shipments and Total PC Market shipment volumes to that value. This is to visualize the amount of variance in the sales numbers on a similar scale. 
The shipping volumes in the overall PC industry show a standard deviation of $27.8$\%  of the mean over the past four years. Notice individual vendors show more variance than the whole industry ($11$\% additional standard deviation). 
Variance in semiconductor revenue over the last 30 years~\cite{semi_revenue} (Figure~\ref{fig:Util_revenue})
illustrates the same issue. Averaged over 30 years, revenue varies over a moving average forecast with a standard deviation of $12$\% of profit, and individual years see deviations of more than $25$\%. The fraction of foundry capacity used~\cite{semi_utilization} also varies greatly due to market volatility.


Supply and demand volatility has only worsened in recent years as  factors of uncertainty in the semiconductor industry are intensifying. On the supply side, climate change is increasing the chance of fires, droughts and winter storms that affect foundries
~\cite{clim_chan_drought_fire}, ~\cite{clim_chan_winter_storm}; rising geopolitical tensions~\cite{t11} ~\cite{t12} are raising the chance of trade embargoes that disconnect entire markets/suppliers from the world system; and increasingly complex designs and lithography techniques are steadily reducing the number of facilities that
can make cutting-edge chips~\cite{t10}. Demand for chips is also increasingly  volatile as they are used in  more and more applications that, with little substitutability among applications (e.g., spike in demand of GPUs when cryptocurrency values surge), cannot be fulfilled by slump or average demand in another application. 
This volatility may have a steep cost, as Table ~\ref{tab:disrupt_cost} shows for some recent unforeseen events.

How can this problem be addressed? 
Given the inherent characteristics of semiconductors and their production (short product life times~\cite{ITIF}, long lead times~\cite{semi_lead_time_long}, etc.) and the fact that modifying semiconductor production takes several years~\cite{fab_takes_long_time}, little can be done in the short term to improve semiconductor supply. While forecasts of demand can help mitigate the effects of demand volatility,  forecasting market behavior is difficult. Figure~\ref{fig:Unc_in_forcast} compares projections from top analytic firms for total yearly semiconductor revenue to actual revenue in 2018--2021
~\cite{Gartner_Semi_Rev_Forc_2Q18, Gartner_Semi_Rev_Forc_2Q19, Gartner_Semi_Rev_Forc_2Q20, Gartner_Semi_Rev_Forc_2Q21, IDC_Semi_Rev_Forc_18, IDC_Semi_Rev_Forc_19, IDC_Semi_Rev_Forc_20, IDC_Semi_Rev_Forc_21},
 showing differences of 30--40 billion dollars per year, despite projections made in the middle of the year. The recent glut of chips in the market \cite{glut_of_chips} also illustrates the failure of current approaches to mitigate supply and demand volatility. 

So if typical ways to address the problem do not work, what can be done? Much  previous work~\cite{Allign_SC_strat_Prod_Unc} suggests finding the right supply chain for the product characteristics. In this paper, we turn that on its head to  explore the right product characteristics (architectures) for the supply-demand chain, and ask the question---{\em What role does chip architecture have in mitigating effects of supply and demand volatility?}

We develop a model (Section~\ref{sec:model_new})
to analyze the impact of different architectural techniques on supply chain costs. Using model parameters derived from real-world data, we study the interactions between architecture and uncertainties in semiconductor supply and demand. We believe our model, rooted in consultation with American and Korean industry practitioners~\cite{begen_supply_2016},  is sufficiently robust to yield several interesting and sometimes counter-intuitive conclusions, since we focus  on comparisons and trends rather than absolute numbers.

We show (Section~\ref{sec:resultsnew}) that a one-to-one chip-to-demand mapping, as is common today, is a significant source of fragility, causing over three-fourth reduction in profits under our estimates of industry uncertainty. We show that one architectural strategy to mitigate this, {\em composition}, has limited efficacy in improving profits beyond the yield savings it offers as it is currently done, showing only a small benefit under supply volatility. A new optimized approach to composition, however,  mitigates up to $33$\% of the losses from demand volatility. We show another architectural solution to eliminate one-to-one chip-to-demand mapping, {\em adaptation}, can mitigate $20$\% of losses from demand uncertainty but is unable to mitigate losses from supply volatility. A new optimized form of adaptation doubles its effectiveness in demand-volatile situations and also provides some benefits in supply-volatile situations. We then explore an additional architectural mechanism--- dispersion---that can  nearly halve losses from supply volatility, and study its interactions with composition and adaptation. Finally, we model how knowledge of market behavior helps improve profits. We find that its benefits are less than most of our proposed interventions, and that these interventions continue to provide benefits when market mechanisms are known. In all, we show that nearly half of the losses from both supply and demand volatility can be mitigated by our  techniques. 

The main contributions of the paper are as follows.
\begin{itemize}
    \item We perform the first study on the role of chip architecture in
       resilience against supply and demand volatility.  
    \item We make a key observation (and demonstrate quantitatively) that a
        one-to-one mapping between produced and demanded chips leads to high
        costs from supply and demand disruptions.
    \item We develop a model that allows us to analyze the impact of different
        architectural techniques on supply chain costs. This is the first such
        model and will be open-sourced as a tool for wider use.
    \item We identify and model two specific architectural strategies to mitigate volatility---composition and adaptation, and a third architecture-supported strategy---dispersion. 
    \item We present  counter-intuitive results based on real-world data about market conditions where these interventions are effective and where they are not, and explore the benefits and drawbacks of combining various interventions.
 \end{itemize}

\begin{table}[]
\small
\centering
\caption{Unforecasted events and resulting loss to foundry} \label{tab:disrupt_cost}
\centering
\begin{adjustbox}{max width=\linewidth}
\begin{tabular}{@{}ll@{}}
\toprule
\multicolumn{1}{c}{\textbf{Event}} & \multicolumn{1}{c}{\textbf{Loss to Foundry (\$)}} \\ \midrule
Power outage at TSMCs legacy process fab (8/2021) ~\cite{t1}                & 28--35M (30--40K wafers)                         \\ \hline
Texas winter storm froze Samsung foundry  (4/2021) ~\cite{t2}                & 286--357M ($\sim$71K wafers)                    \\\hline
Texas winter storm froze NXP foundry (4/2021) ~\cite{t3}                    & $\sim$100M                                      \\\hline
Fire at Renesas facility in Naka Japan (3/2021) ~\cite{t4}                  & 171M                                         \\\hline
Global Foundries quits on 7nm,   AMD shifts to TSMC (8/2019) ~\cite{t5,t6} & $\sim$4.255B loss just from AMD             \\\hline
Current chip shortage due to surge in demand (2020) ~\cite{t7}           & 60B    \\\hline
China-US trade war causes TSMC to lose HWAWEI sales (2020) ~\cite{t8}  & 13\% TSMC's revenue $\sim$= 455M              \\\hline
Fire at Asahi Kasei Foundry (2/2021) ~\cite{t9}                             & 100M                 \\ \bottomrule
\end{tabular}
\end{adjustbox}
\end{table}

\section{Volatility-resilient chip architectures}
This section illustrates a problem that may be due to product design and introduces \emph{composition} and \emph{adaptation} as two architectural strategies that potentially improve resilience to volatility in supply and demand. 

\begin{figure}
    \centering
    \includegraphics[width=0.5\linewidth]{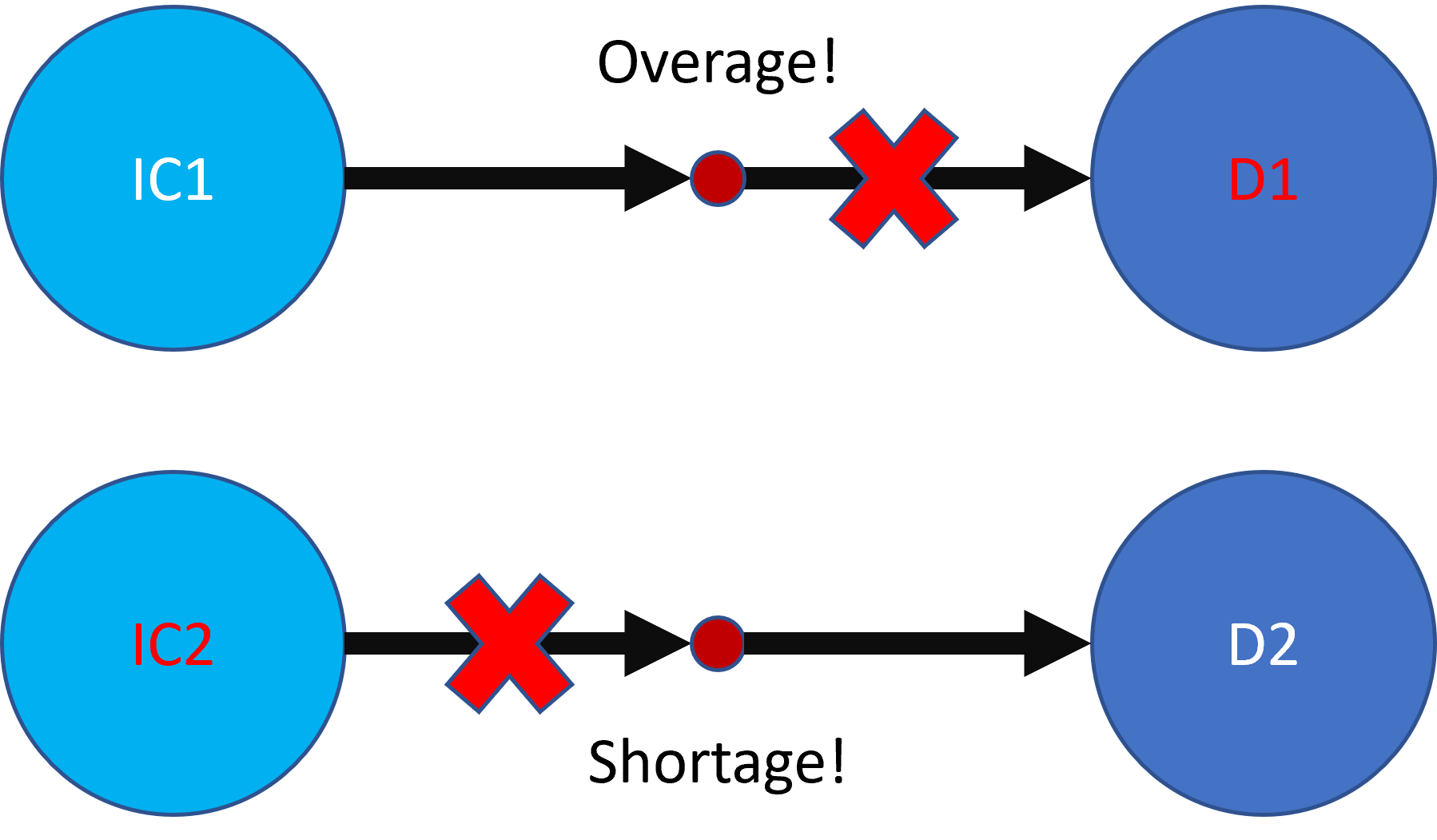}
    \caption{\centering Fragility of single-demand paths: IC1 wasted when D1 falters, D2 left un-met when IC2 falters}
    \label{fig::one-to-one}
\end{figure}

\subsection{Fragility of single-demand paths}
Consider single-demand paths, where one chip can satisfy only one demand; such paths may still occur in spite of the wide prevalence of binning when decisions around binning are made only with the goal of maximizing yield.
Figure \ref{fig::one-to-one} depicts the clear problems with this approach---when there is only one supply for a demand, the supply input components (IC1) are wasted when demand (D1) falters, and the demand (D2) is not met when supply input component IC2 falters. The proposed architecture strategies below attempt to spread risk for different supplies and demands across products.

\begin{figure}[ht]
    \captionsetup[subfloat]{farskip=0pt,captionskip=1pt,nearskip=-2pt}
    \centering
    \subfloat[\centering \small Adaptation is a one-to-many relation]{
        {\includegraphics[width=0.4\linewidth]{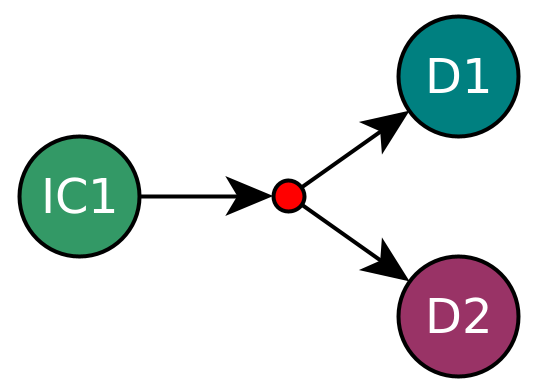}}
        \label{fig:adaptation_graph}
    }
    \subfloat[\centering \small Adaptation can create alternate paths to the
    same demand]{
        {\includegraphics[width=0.4\linewidth]{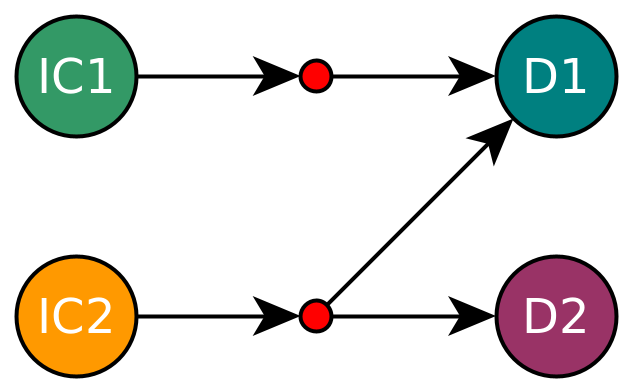}}
        \label{fig:ada-multi-path}
    }
    \caption{Benefits of Adaptation}
    \label{fig:adaptation-mappings}
\end{figure}

\subsection{Adaptation}

The adaptation strategy involves architecting and producing a chip that can be modified (“adapted”) with low/no overhead, to satisfy multiple types of customer demands. Examples of architectural realizations of adaptation include disabling cores~\cite{harding_2018} or reducing frequencies to sell a chip at a different performance tier~\cite{binning}. Multi-ISA designs~\cite{venkat2014harnessing} could also be an architectural realization of adaptation---different demands may make use of the different ISAs the chip supports. Theoretically, adaptation may be able to provide benefits because it delays product differentiation closer to when demand is known. Figure \ref{fig:adaptation-mappings} depicts adaptation as a one (chip) to many (demands) mapping. Suppose a firm produces an input component IC1-based saleable product which satisfies demand D1, but can be adapted to also satisfy demand D2 (Figure \ref{fig:adaptation_graph}), and there is a spike in demand for D2 - in that case the firm can adapt a higher proportion of the IC1 it produces to satisfy D2. Additionally, adaptation can create robustness by enabling alternate paths to the same demand (figure \ref{fig:ada-multi-path}) - demand D1 is primarily satisfied by IC1, but if IC2 can be adapted to also be able to satisfy D2, there is redundancy if the supply for D1 if IC1 cuts out. While theoretically adaptation can help mitigate affects of uncertainty, we explore exactly how this may happen in different market volatility situations and the extent of its benefits, in the results section.

\begin{figure}[ht]
    \captionsetup[subfloat]{farskip=0pt,captionskip=1pt,nearskip=0pt}
    \centering
    \subfloat[\centering \small Composition mitigates demand
    uncertainty risks]{
        {\includegraphics[width=0.5\linewidth]{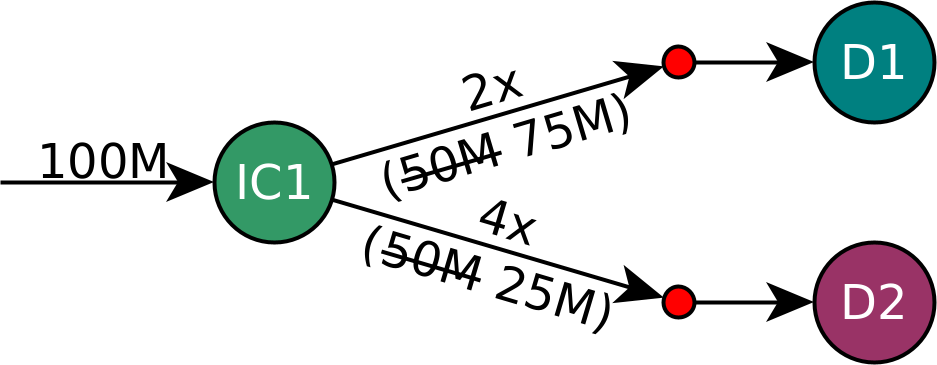}}
        \label{fig:composition_graph2}
    }
    \subfloat[\centering \small Composition mitigates supply
    uncertainty risks]{
        {\includegraphics[width=0.4\linewidth]{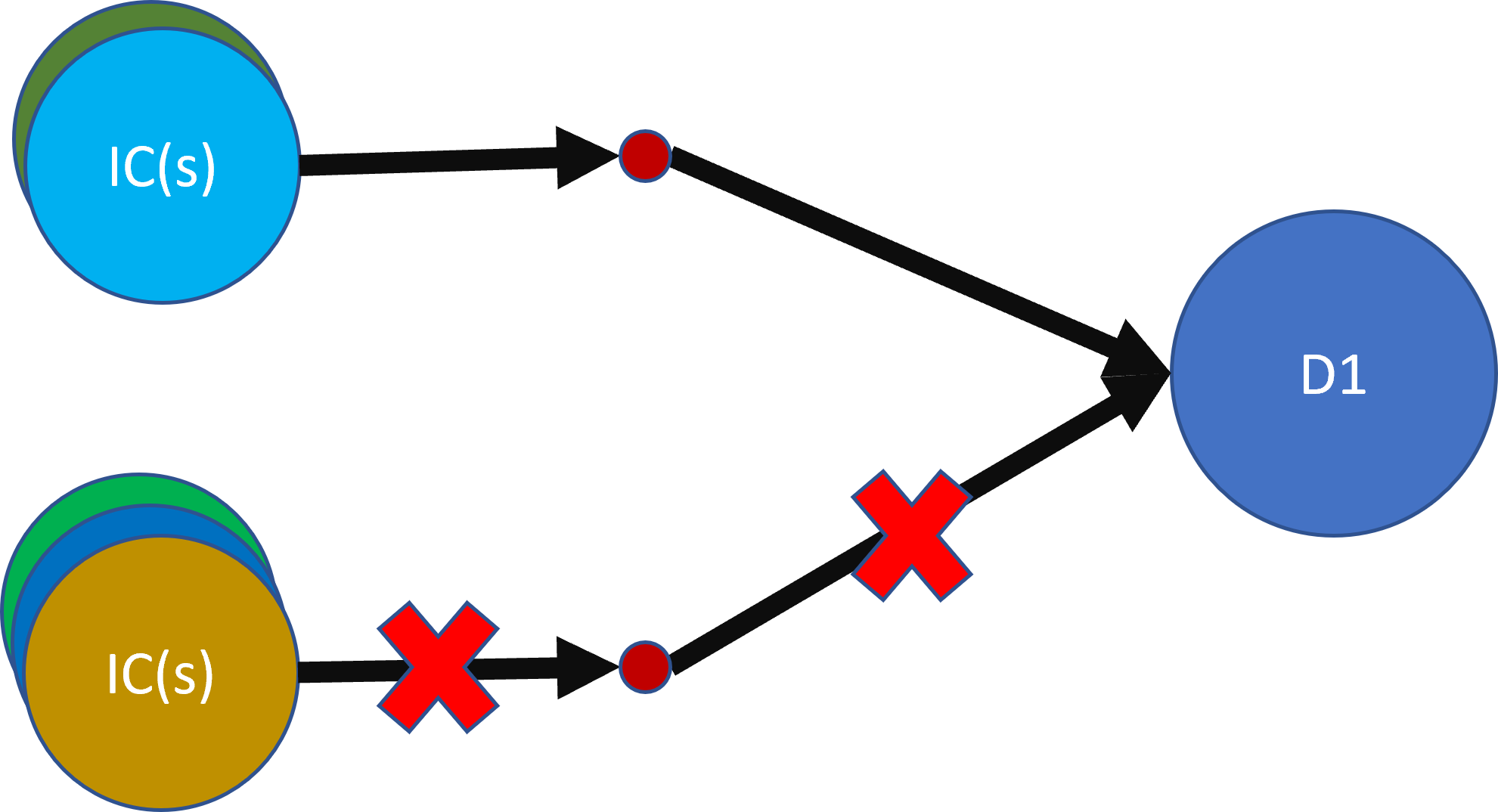}}
        \label{fig:composition_graph3}
    }
    \caption{Benefits of Composition}
    \label{fig:compsition-mappings}
\end{figure}
\subsection{Composition}

The composition strategy involves designing and producing sub-components of a chip. These sub-components can then be combined in different quantities to create packaged chips that cater to different demands. Figure \ref{fig:compsition-mappings} depicts composition as a many (sub-components) to one (chip) mapping. An example of a current architectural realization of composition is chiplet-based architectures~\cite{amdepycandryzen}, where chiplets can be interconnected (“composed”) using an interposer to make a larger chip. Theoretically, composition can provide benefits when a) there is an intersection in the sub-components used to produce different chips and b) there are multiple (independent) sets of sub-components that can be composed to make chips that satisfy the same demand. As depicted in Figure \ref{fig:composition_graph2}, (a) helps by allowing the chip firm to decide what packaged chip a sub-component will go toward manufacturing, closer to when it is sold. This is effectively delaying product differentiation much like adaptation. Figure \ref{fig:composition_graph3} depicts (b) where if the sub-components in path 1 needed to produce the demand goods are suddenly unavailable, the demand can still be satisfied through the sub-components in path 2. Thus helping alleviate the effects of supply volatility. While these theoretical characteristics of composition can potentially provide resilience to uncertainty, a deeper dive in the results section shows a lot more interesting behavior in terms of the ways and extents to which composition can address volatile markets.

\section{The Model}
\label{sec:model_new}

To understand how different architectural techniques perform in an uncertain market,
we develop a model that computes the profit realized by a chip
firm with and without disruptions to demand and supply. 
We build on the cost model of a generic supply chain developed by
Begen {\em et al.}~\cite{begen_supply_2016}; this model can estimate the impact
of supply and demand uncertainties on supply chain costs. Crucially, the model
was developed in consultation with  industry experts at Intel and
"a large Korean consumer electronics firm". We augment the model to enable
compatibility with commercial stochastic program solvers and the ability
to model different computer architecture techniques. Since we are less
interested in the absolute numbers produced by the model than the (especially
comparative) trends and analyses generated by the model, we believe that this
model has sufficient fidelity for the nature of the conclusions we derive
(Section~\ref{sec:resultsnew}).

In this paper, we are primarily concerned with the effects of market uncertainty on chip manufacturers. Hence the model we develop 
optimizes for the highest mean total profit that the firm realizes, given the market conditions. In the results section (Section~\ref{sec:resultsnew}) 
we use this as our primary metric to assess the effectiveness of the proposed architectural techniques. The high-level equation for profit is given in Eq. \eqref{eq:tc}, with the following sections explaining each of the terms in the equation and how uncertainty is modeled into this equation.   

\begin{equation}
    P = TC_{ben} - TC_{prod} - TC_{map} - TC_{short}  \label{eq:tc} 
\end{equation}

\begin{table}
\small
    \caption{Model Parameters}
    \label{tab:ComPar}
    \centering
    \begin{tabular}{|c|c|} \hline
        \textbf{Parameters} & \textbf{Description}  \\ 
        \hline
        $\uc{}$ & Per-Unit Cost\\
        $\yield{}$ & Yield Rate \\
        $\nre{}$ & NRE Cost\\
        $\uben{}$ & Per-Unit Benefit\\
        $\usc{}$ & Shortage Cost \\
        $\based{}$ & Base Demand \\
        $\vec{\gamma}$ & Mapping Cost \\
        $\vec{M}$ & Mapping Functions \\ \hline
        \textbf{Random Variables} &  \\\hline
        $\zs{}$ & Supply Uncertainty \\
        $\zd{}$ & Demand Uncertainty \\ \hline
        \textbf{Decision Variables} &  \\\hline
        $\ord{}$ & Ordered \\
        $\quant{}$ & Order Quantity\\ 
        $\mappingquant{}$ & Used Mappings \\ \hline
    \end{tabular}
\end{table}

Table \ref{tab:ComPar} contains a list of parameters we pass to the model. Further, it contains a list of the random and decision variables. Random variables $Z^s$ (Supply Uncertainty) and $Z^d$ (Demand Uncertainty) are used simulate uncertainty in the market. We use these random variables to scale the supplied quantity and amount demanded. Finally, the decision variables are values the model (the firm) can tweak in order to maximize profit.

\subsection{Profit Calculation}
\subsubsection{Cost of Production}
To satisfy market demand for chips, a chip firm must first decide what types of chips it puts engineering effort into developing. $\ord$ represents this decision. It is a binary vector where $o_i$ is set to one if and only if the firm decides to design a chip of type `i'. For each chip of type `i' that the firm chooses to develop, it incurs a cost $n_i$ for the non-recurring engineering (NRE) costs involved in designing chips of type `i'.

Once chips are designed, a chip firm must produce them to sell to customers, so the firm decides to order, from a semiconductor foundry, various quantities of the different chips it designed. 
Let $\quant$ denote this order quantity decision, where $q_{i}$
represents the number of chips of type `i' the firm ordered. In 
our experiments, these can be 16-core, 8-core or 4-core chip(let)s. 
Note that $o_{i}$ is zero if and only if $q_{i}$ is zero; in all other cases, it is one.  
Once chip order quantities are decided, then based on supply uncertainty, $\zs{}$, the firm receives more (or fewer) chips than it ordered. We represent this received quantity as $\crecv{}$ and it is the order quantity scaled by supply uncertainty:
\begin{equation*}
    \crecv = \hadprod{q}{Z^s}.
\end{equation*}
After the chips are received, we use $\yield{}$, the average yield for each of the chips, to realize what fraction of these received chips is usable. The number of usable chips, $\cobt{}$, is given by
\begin{equation*}
    \cobt = \hadprodnv{\crecv}{\yield}.
\end{equation*}
Each of the chips has a per-unit cost $\ucost$. The cost of production incurred by the firm for a particular type of chip is the sum of the unit costs of the received chips, and the non-recurring engineering (NRE) costs incurred in the design. Total production cost for the firm is a sum across all the types of chips:
\begin{equation} \label{eq:tc_prod}
    TC_{prod} = \crecv{}\cdot{} \uc{} +\ \ord{}\cdot{}\nre{}. 
\end{equation}

\subsubsection{Cost of Mapping}

Once the number of chips obtained is known, the firm can decide how to use these 
chips to maximize profit. The model takes as input a list of mapping functions $\vec{M}$. 
Each mapping function $M_{j}$ describes a set of inputs $M^{inp}_j$ and outputs $M^{out}_j$:

\begin{equation*}
    M_j = (M^{inp}_j \ ,\ M^{out}_j).
\end{equation*}

Take the baseline case where the firm has a single path from demand to supply in a market for 16-core, 8-core and 4-core chips. In this case, we specify three mapping functions.

\begin{align*}
    M_1 &= \text{(16 core, 16 core)} \\
    M_2 &= \text{(8 core, 8 core)} \\
    M_3 &= \text{(4 core, 4 core)}.
\end{align*}

In the case of composition and adaptation, additional mapping functions are added to allow for one-to-many and many-to-one mappings between obtained and sold chips. For example, during adaptation, the firm has the ability to re-purpose 16-core chips as 8-core chips, so mapping function $M_x$ = (16-core, 8-core) is added to the model. 

The decision variable $\mappingquant{}$ is used to determine how each mapping function is used, where $U_j$ denotes the number of $M_j$ mappings used.  The number of used chips $\cused{}$ is given by 

\begin{equation*}
    \cused{} = \vec{U}  \cdot \vec{M^{inp}}
\end{equation*}

Since the number of chips used cannot be more than the number obtained after supply uncertainty and yield losses, the following condition is added to the model

\begin{equation*}
    \cusednv^i \leq \cobtnv^i~\forall\ i.
\end{equation*}

Using $\vec{U}$ and the mapping output $\vec{M^{out}}$, the model decides the number of chips obtained that can be built for potential sale. Let $\cbuilt{}$ be the number of chips the model decides to `build': 

\begin{equation*}
    \cbuilt = \mappingquant{}\cdot\vec{M^{out}}.
\end{equation*}

Finally, each mapping function has a cost associated with it. This cost may, for example, represent interposer costs in the case of composition mappings. This cost is an input parameter
given by $\vec{\gamma}$, such that $\gamma_j$ is the mapping cost associated with $M_j$. Then the total mapping cost incurred by the firm is given by 

\begin{equation} \label{eq:tc_map}
    TC_{map} = \vec{U}\cdot\vec{\gamma}.
\end{equation}

\subsubsection{Revenue}

For all evaluations, the model simulates a market demand for 16-core, 8-core and 4-core chips. Each of the demanded chips `i' has a base-demand of value $b_i$, which is a model 
parameter. The actual demand for the chips, $\cdemand{}$, is given by 

\begin{equation*}
    \cdemand{} = \hadprodnv{\based{}}{\zd}. 
\end{equation*}

where $Z^d_i$ is the demand uncertainty for chip of type 'i'. The number of chips sold is determined by the $\cdemand{}$ and $\cbuilt{}$ - the firm cannot sell more chips than it has built and it cannot sell more chips than the amount demanded.
The model takes the minimum of the two and uses that as the number of chips sold, $\csold{}$, given by 

\begin{equation*}
    \csold = min(\cdemand{} , \ \cbuilt{}).
\end{equation*}

Chips of type 'i' are each sold at a price $u_{ben}^i$ and hence the total revenue is given by

\begin{equation}\label{eq:tc_ben}
    TC_{ben} = \uben{} \cdot\ \csold{}.
\end{equation}

\subsubsection{Shortage Cost}
If the firm builds fewer chips than what were demanded, it incurs an additional cost called shortage cost. In addition to the
opportunity cost of missed potential sales, there is an additional shortage
cost to the firm for each chip it didn't deliver because it didn't satisfy the
customer's requirement (concretely, this could be in the form of lost back-orders, lower
future order quantities etc.). The per unit shortage cost for a chip 'i' is given by $u_{sc}^i$ and in all the experiments, the shortage cost of chip 'i' is equal to its unit benefit $u_{ben}^i$. Thus, the total shortage cost incurred by the the firm is 

\begin{equation} \label{eq:tc_short}
    TC_{short} = \usc{} \cdot\ (\cdemand{} - \csold{}).
\end{equation}

In this way, equation (\ref{eq:tc}) represents the net profit the firm makes by combining equations (\ref{eq:tc_prod}), (\ref{eq:tc_map}), (\ref{eq:tc_ben}) and (\ref{eq:tc_short}). 

\subsection{Modeling Uncertainty}
\label{sec:modeling_uncertainty}
For all our experiments, we model supply and demand uncertainties as normal distributions with a mean of one and vary standard deviation to simulate uncertainty. In most experiments, all elements in \(\zs{}\) are sampled from a single normal distribution, i.e, all orders have the same realised surplus or deficit. In some cases, it is a combination of multiple distributions, for example in the case of dispersion. Unlike in the case of supply uncertainty, demand uncertainty, \(\zd{}\), is unique to each kind of chip and this is consistent across all experiments.
For each evaluation, we sample multiple values for $\zd{}$ and $\zs{}$. Then the decision variables are optimized to maximize $\E{P}$ given by

\begin{small}
\begin{equation*} 
    \E{P} = \E{TC_{ben}} - \E{TC_{prod}} - \E{TC_{map}} - \E{TC_{short}}.
\end{equation*}
\end{small}

\section{Methodology}

\subsection{Simulation Framework}
To accurately assess the impact of architectural techniques on supply
and demand uncertainty, we identify \textit{optimal} chip order ($\quant{}$ \& $\ord{}$) and
chip mapping policies ($\vec{U}$) that maximize expected total profit, $\E{P}$. This is done so results under different interventions and market conditions are comparable. 
Note that some decision variables need to be determined independent of some random variables. For example, $\quant{}$ and $\ord{}$ need to be decided before demand and supply are known. Also $Z^d$ and $Z^s$ are probability distributions, not numeric parameters. So, in some cases, a single decision is made for all the samples in the distribution (exception to this are detailed later). 

\begin{table}[ht]
\caption{\centering Staging for Decision Variables and Recourse Parameters}
\label{tab:recourse}
\centering
\begin{tabular}{@{}ccc@{}}
\toprule
\multicolumn{1}{l}{\textbf{Decision Variable}} & \multicolumn{1}{l}{\textbf{Decision Stage}} \\ \midrule
$q$   & 1     \\
$o$ & 1  \\
$U$   & 2 or 3    \\ \midrule
\textbf{Random Variable} & \textbf{Recourse Stage}  \\ \midrule
$Z^s$  & 2      \\
$Z^d$ & 2 or 3 \\ \bottomrule
\end{tabular}
\end{table}

Since this optimization problem is a stochastic mixed-integer non-linear
program, we use a multistage stochastic program with recourse
(MSPR)~\cite{mspr,mspr2} to solve it.
The idea of an MSPR is to sequence the decision variables across multiple
temporal stages, with latter stages having `recourse' to information that was
not available in earlier stages. 
We perform simulations for different permutations of decision-making stages and recourse. These are summarized in Table \ref{tab:recourse}. The value of order quantity $\quant{}$ and decision to order $\ord{}$ are always determined in stage 1, without recourse to supply and demand. Mapping $\mappingquant{}$ is determined either in stage 2 once supply is known or in stage 3 once both demand and supply are known. We offer this flexibility for mapping since, for example, the decision to sell a 16-core processor as an 8-core processor
can be made very late in the production process, since disabling cores can
be done via firmware. So this can be done once demand is realized. 

We use the global solver in the commercial mathematical
programming software Lingo~\cite{lingo} for finding globally optimal chip orders and mappings. We generate Lingo models using a Rust library which transforms
collections of produced goods, demanded goods, mappings, and additional
constraints into Lingo models.

\subsection{Model Parameters}


For a study like this, it is critical that model parameters represent reality.
Publicly available
data~\cite{lee_semiconductors_2022,noauthor_results_2022,noauthor_coronavirus_nodate, noauthor_gartner_nodate}
suggests that most fluctuations of semiconductor products remain within $\pm 50\%$ of
expectation. Additionally, our analysis of semiconductor revenue over the past 30 years (Figure \ref{fig:Util_revenue}) shows up to a 15\% standard deviation in profit over a 10 year period. Our analysis of the PC market over the last four years (Figure~\ref{fig:PC_mar_fluct}) shows a 27\% standard deviation in the quantity of PCs shipped, and that individual vendors around a 11\% higher standard deviation in profit with respect to the industry, or about a 30\% standard deviation. 
As such, our experiments are run between the standard deviation values for the semiconductor industry and for individual vendors in the PC industry, with some additional tests on either side of that range.

We performed several case studies centered around \SI{22}{\nano\meter} 16, 8,
and 4-core dies. To estimate their area, we scaled the area for a 32-core
monolithic-chip from~\cite{amdnext} based on the number of cores and then
scaled it further to \SI{22}{\nano\meter}.
To estimate the yield, $\yield{}$, of these dies, we used the model in a previous study from AMD~\cite{amdpaper}
($Yield = (1+(D_0/n \cdot A_{crit}) / \alpha)^{-\alpha}$) assuming a
\SI{300}{\milli\meter} wafer, 12 metal layers,  $Frac_{crit}$ (wire)  of 0.2625,
$Frac_{crit}$ (logic) of 0.75, and clustering factor $\alpha$ of 1.

To estimate the unit cost, $\ucost$, of each of the dies, we first estimate the recurring engineering (RE) cost for a given order as 
$RE\_cost = (order/N_{dies})*RE\_wafer\_cost\_logic$, where $RE\_wafer\_cost\_logic = 24300$ was taken 
from~\cite{tvlsi} (Table IV). The cost estimates (normalized w.r.t \SI{45}{\nano\meter} wafer cost) are largely based on published industrial data. 
To estimate the NRE cost, $\nre{}$,  we again used cost numbers 
from~\cite{tvlsi} (Table IV). NRE cost has two parts, 1) area dependent NRE design cost and 2) area independent NRE mask set cost. We scale the design cost with our core area and add one mask set cost per design to estimate the total NRE cost. Interposer costs are derived the same  way using numbers 
from~\cite{tvlsi} (Table IV).  For validation, we calculated the cost of a 32-core monolithic chip and 32-core chip composed of 4 8-core
chiplets through our model. We found the 32-core chip built from composing 8-core chiplets was 1.71× cheaper. 
For comparison, \cite{amdnext} claims that their 32-core chip composed using 
4x8-core chiplets is $1.69\times$ cheaper to 
build than their 32-core monolithic-chip.  

For each demanded chip, we assume a base demand ($\based{}$) of $10^8$ chips. Unit benefits were computed as follows $u_{ben} = 2\left(u_{cost} + \frac{n}{b}\right)$ (assuming a 50\% gross margin \cite{amd_margin,intc_margin}). Due to the importance of time-to-market, unit
shortage cost, $\usc$, is equal to unit benefit $\uben$.

We use mapping functions ($\vec{M}$) and the mapping decision variable ($\mappingquant{}$) to realize our architectural techniques - composition and adaptation. Each mapping ${M_j}$ has a mapping cost $\gamma_j$ associated with it. 
For composition, we primarily explore monolithic chip vs chiplet based designs. Chiplet based designs have a variable interposer costs. Hence, for any mapping that performs composition, $\gamma$ takes a value for an interposer of area equal to the overall produced chip area. 

\begin{figure}
    \centering
    \includegraphics[width=1\linewidth]{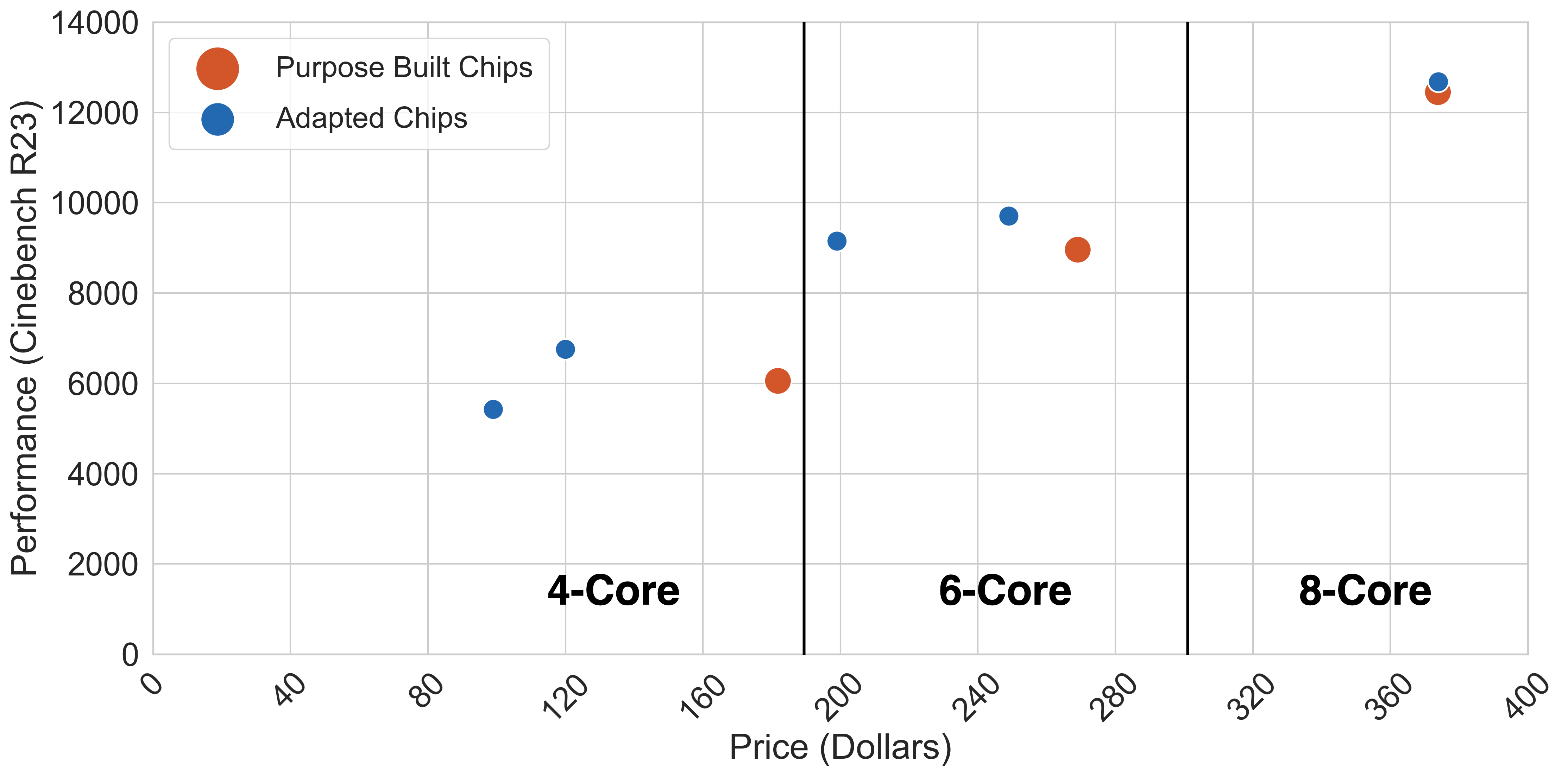}
    \caption{\centering Price and Performance for Adapted vs Purpose Built Chips}
    \label{fig:adaptation_cost}
\end{figure}

Our adaptation technique assumes mappings from larger chips to smaller chips. For example, a 16-core or 8-core chip can be rebadged and sold as an 8-core or 4-core chip, given market constraints. To estimate adaptation cost, we looked at real world data. Figure~\ref{fig:adaptation_cost}  shows the relationship of inflation-adjusted price and performance for purpose-built vs adapted chips for several consumer chips from Intel and AMD. Prices for chips with the same core count are roughly the same and also show similar performance (Cinebench score \cite{4core1,4core2,8core,6core1,6core2}). Hence, we allow unit benefit of adapted chips to be the same as monolithic chips built for that purpose and the mapping cost of this intervention is zero. 

\subsection{Experiment Configurations}
\label{sec:config}
In Section \ref{sec:resultsnew}, we simulate a market for 4-core, 8-core, and 16-core chips using the model and methodology above. Unless otherwise mentioned,  we use a single normal distribution to sample supply uncertainty for all the chips, $\zs$, and three independent normal distributions to sample demand uncertainty for each of the chips, $\zd$. For our baseline case, we study a rigid one-to-one mapping between produced goods and demanded goods. 

For our architectural intervention techniques, composition and adaptation, we introduce new mappings. In composition, we allow the firm to produce 16-core chips using two 8-core, four 4-core, and a combination of 8-core and 4-core chiplets. We also add a mapping to allow 8-core chips to be built using two 4-core chiplets. The mapping cost (interposer cost) for the composed 8-core chip is half that of a composed 16-core chip since the size of the interposer used is halved. In adaptation, we allow manufactured 16-core chips to be repurposed and sold as 8-core or 4-core chips, and manufactured 8-core chips to be sold as 4-core chips. 


\section{Results}
\label{sec:resultsnew}




Here we present results on the mean and standard deviation of total profit  under several regimes of uncertainty to understand the benefits and drawbacks of different architectural interventions. The mean profit for the firm is calculated across various samples of demand and supply. And the standard deviation in profit highlights the variance in profit seen across these samples. We present results as line plots, separately for mean profit and for standard deviation of expected profits. To isolate and better and understand the benefits of the various architectural interventions, we simulate market conditions with uncertainty only in the supply of goods, only in the demand for goods, and with both. We study the mappings used by the optimizer to justify the benefits and drawbacks of the various experiments. An important metric that we consider is lambda ($\lambda$), which is the percentage of the loss incurred due to uncertainty that is made up by the intervention. Lambda is calculated as

\begin{small}
\begin{align*}
    \lambda&=\frac{\text{(intervention mean profit  - baseline mean profit)*100}}{\text{baseline mean profit with 0 uncertainty - baseline mean profit}}\\
\end{align*}
\end{small}

\subsubsection{Fragility of single-demand paths}

\begin{figure}[ht]
    \centering
    \captionsetup{skip=0pt}
    \includegraphics[width=1\linewidth]{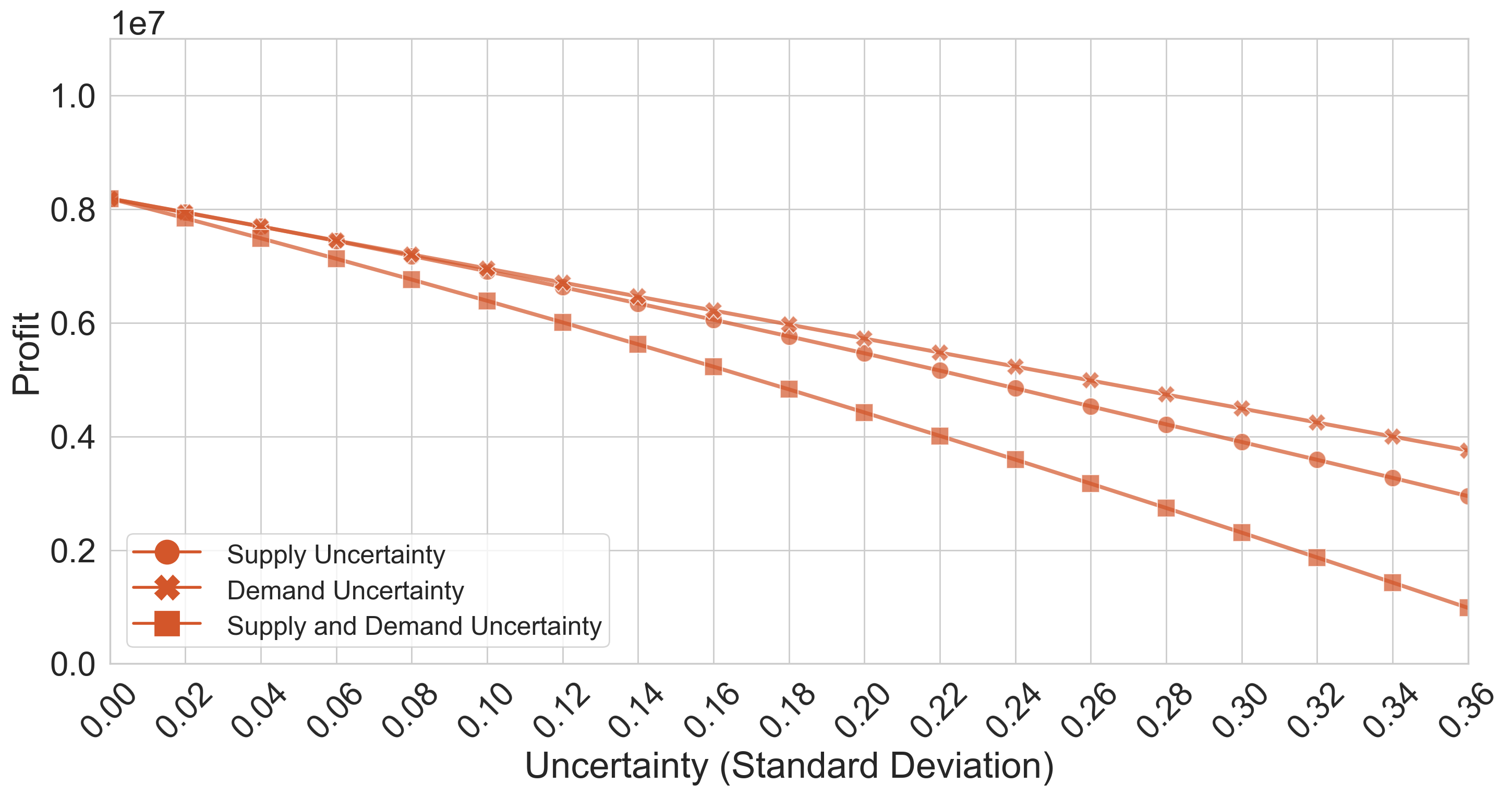}
    \caption{\centering Mean Profit for Baseline Single-Demand Path}
    \label{fig:Baseline_profit}
\end{figure}

\begin{figure}[ht]
    \centering
    \captionsetup{skip=0pt}
    \includegraphics[width=1\linewidth]{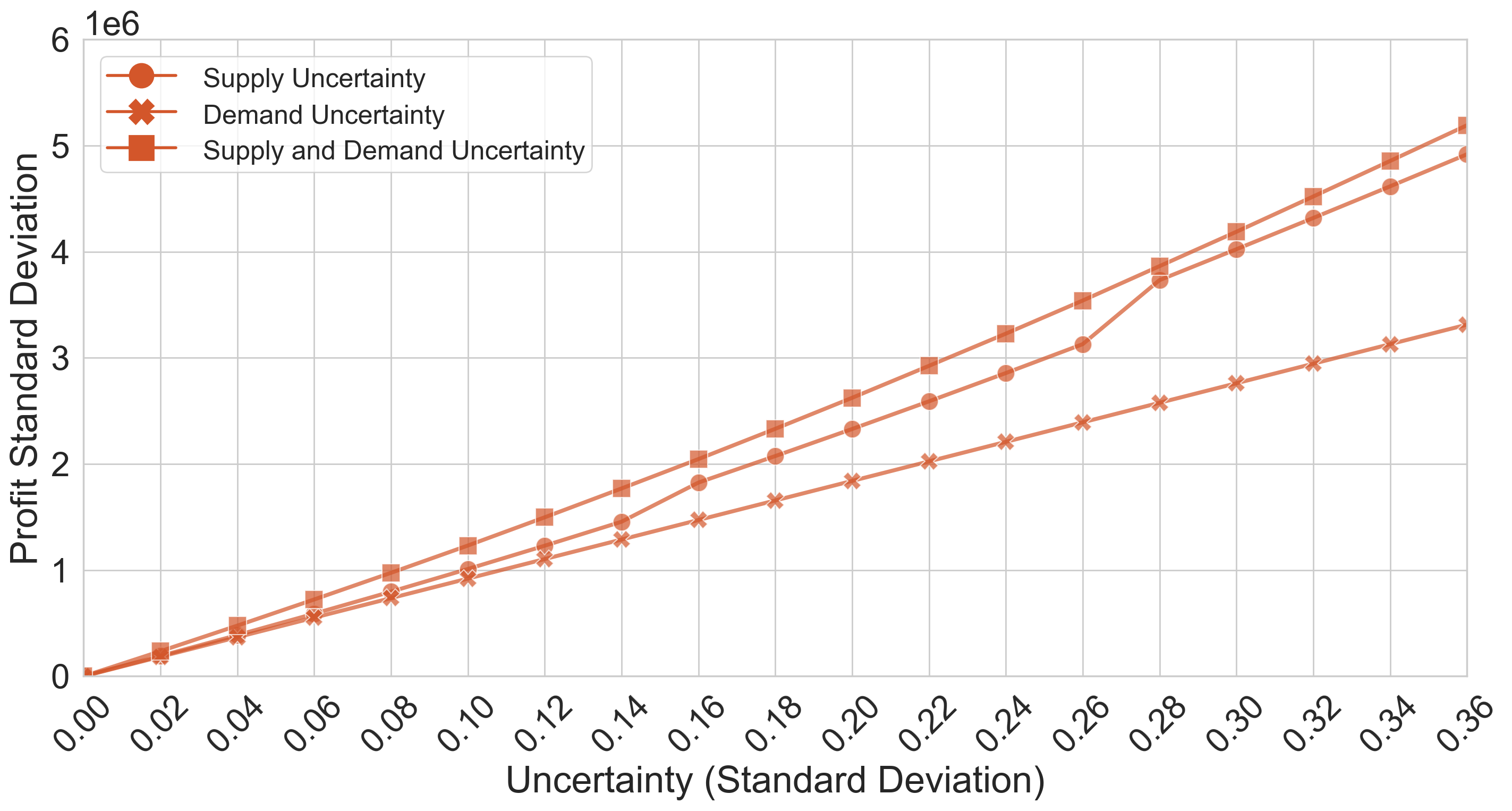}
    \caption{\centering Standard Deviation of Profit for Baseline Single-Demand Path}
    \label{fig:Baseline_std}
\end{figure}

Our baseline results demonstrate the fragility of having only a single path between produced and demanded chips. The baseline models supply and demand for three different chips (4, 8 and 16-core) and each chip is sold exactly as it is produced. \\

The results for the baseline (Figure ~\ref{fig:Baseline_profit}) show that at the highest level of uncertainty (standard deviation of 0.36) the firm sees a \textbf{93\%, 63\%, and 54\% drop in mean profit} for supply+demand uncertainty, supply uncertainty and demand uncertainty, respectively. 
This drop is attributable to the inability of the supply chain to cope with uncertainty. 
Not only is there a significant decrease in average profit, there is also an increase in the range of profits the company could see (Figure~\ref{fig:Baseline_std}). At high levels of variance, the \textbf{standard deviation of profit is almost equal to the mean profit when there is no variance}. 
Clearly, this level of reduced expected profit and variance in profit is undesirable.

\subsection{Architectural Strategies}

\subsubsection{Composition}

\begin{figure}[ht]
    \centering
    \captionsetup{skip=0pt}
    \includegraphics[width=1\linewidth]{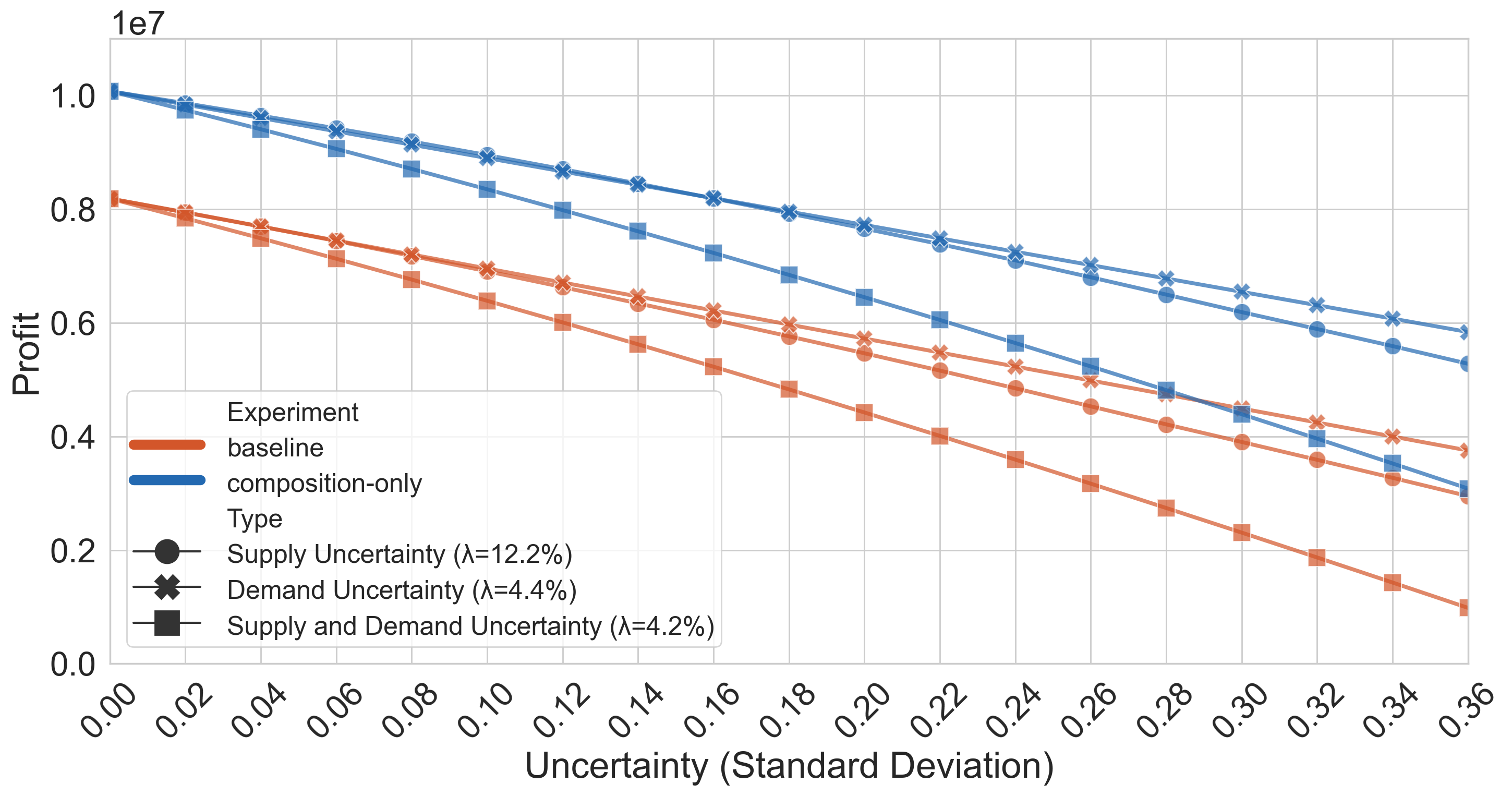}
    \caption{\centering Mean Profit with Composition}
    \label{fig::composition-only_profit}
\end{figure}

\begin{figure}[ht]
    \centering
    \captionsetup{skip=0pt}
    \includegraphics[width=1\linewidth]{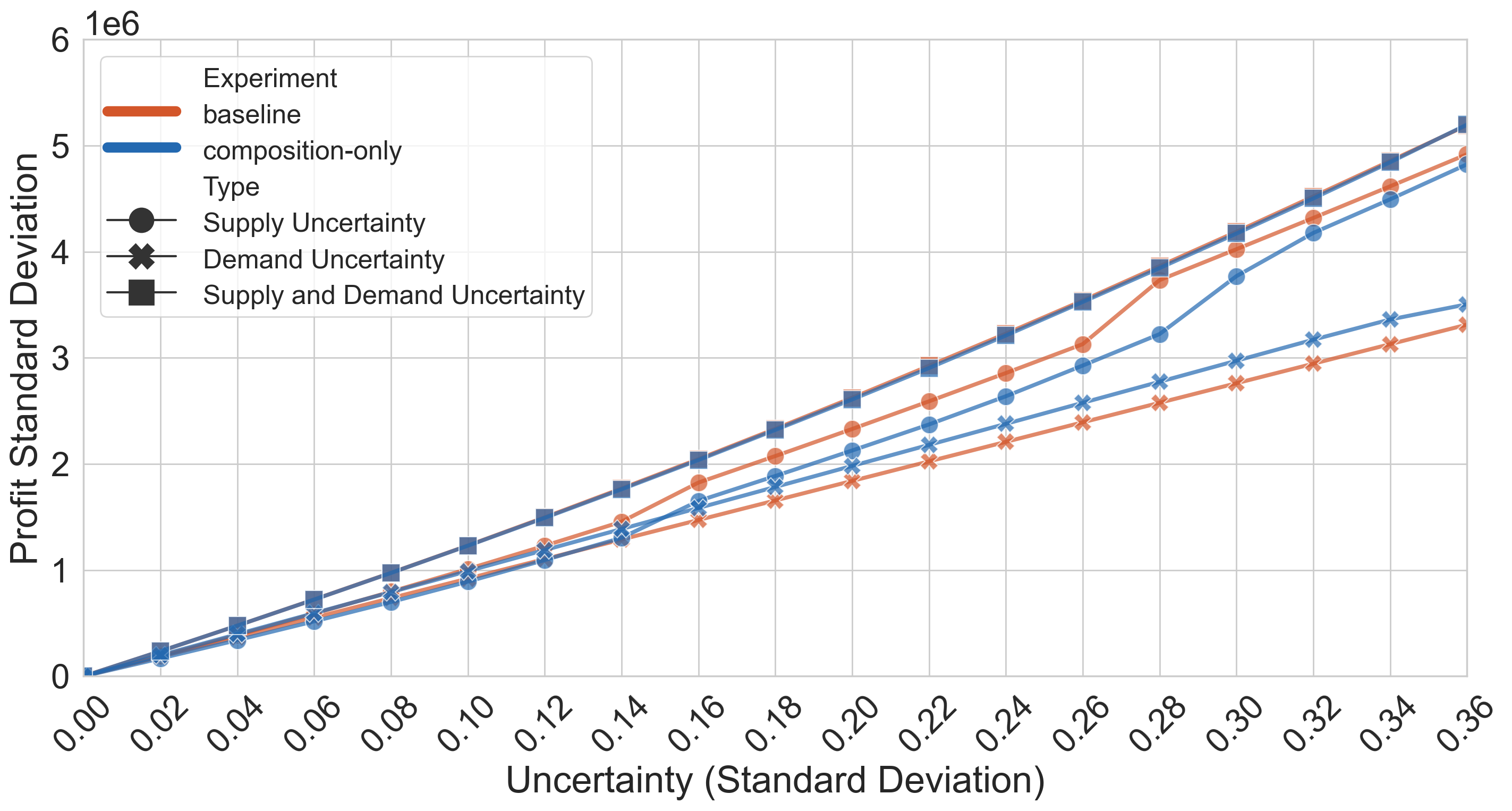}
    \caption{\centering Standard Deviation of Profit with Composition}
    \label{fig::composition-only_std}
\end{figure}

As discussed earlier, composition can enhance resilience by enabling multiple paths to produce the same good. Like the baseline, there are 4, 8, and 16-core goods produced, but here these can be composed in any way possible to make 4, 8, and 16-core packaged chips. An added interposer cost is incurred for every chip produced using composition. Here we use the three-stage model where supply uncertainty is realized first, and then demand uncertainty is resolved after chips are mapped (packaged) using $\mappingquant{}$. These assumptions and configurations are made keeping in mind how composition is carried out currently - chiplets are sourced from the same supplier, and chiplet-based chips end up being sold several months after they are packaged, so demand is not known at the time of packaging.

Results for mean profit using composition are presented in Figure~\ref{fig::composition-only_profit}. We see that mean profit is 25\% higher with composition compared to the baseline when there is no variance. All of this improvement is attributable to yield benefits of composition, however.
Indeed, an analysis of mappings used ($\mappingquant{}$), shows that when there is no uncertainty in the market, the model orders only 4-core chips and makes 16-core and 8-core chips out of these. 
After factoring out the improvement due to yield benefit, we see $\lambda$ is only 12.2\% when there is supply uncertainty, 4.2\% when there is supply and demand uncertainty, and 4.4\% when there is only demand uncertainty. The absolute increase in profit (between composition and baseline) remains almost steady when demand uncertainty is present across the range of uncertainties studied. There is a small improvement when there is only supply uncertainty because the the limited supply of chips under supply constraints is directed to satisfy the demand with the greatest profit (the 16-core part). This suggests that \textbf{composition by itself {\em hardly} improves profits} any more than it does at 0 uncertainty.

Results of profit variance using composition are presented in Figure \ref{fig::composition-only_std}. 
The results show that \textbf{composition does have benefits to profit variance}, especially in regimes of exclusive supply uncertainty. 
Profit-variance reduction is negligible in markets with only demand uncertainty. 

\subsubsection{Adaptation}

\begin{figure}[ht]
    \centering
    \captionsetup{skip=0pt}
    \includegraphics[width=1\linewidth]{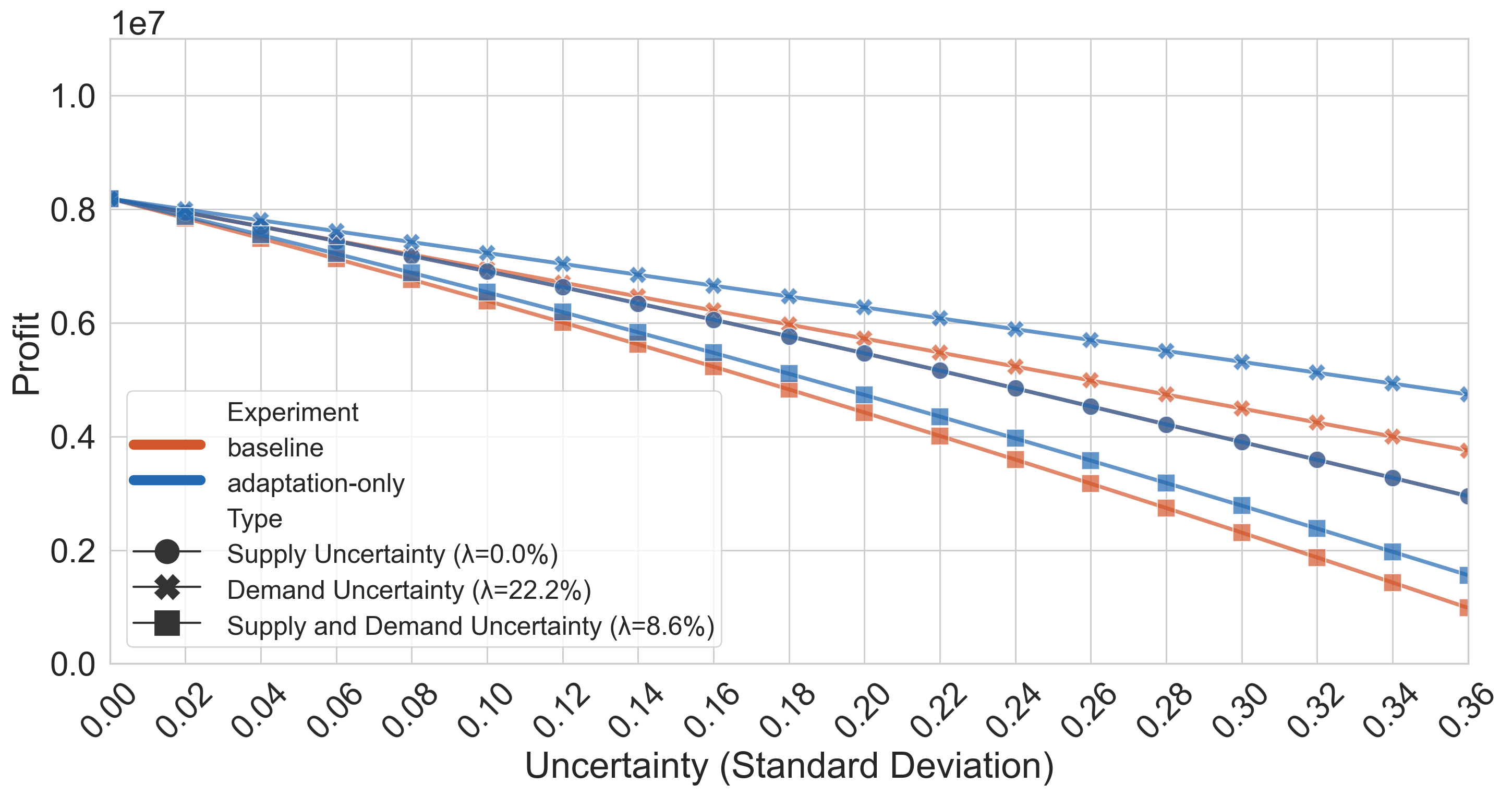}
    \caption{\centering Mean Profit with Adaptation}
    \label{fig:adaptation-only_profit}
\end{figure}

\begin{figure}[ht]
    \centering
    \captionsetup{skip=0pt}
    \includegraphics[width=1\linewidth]{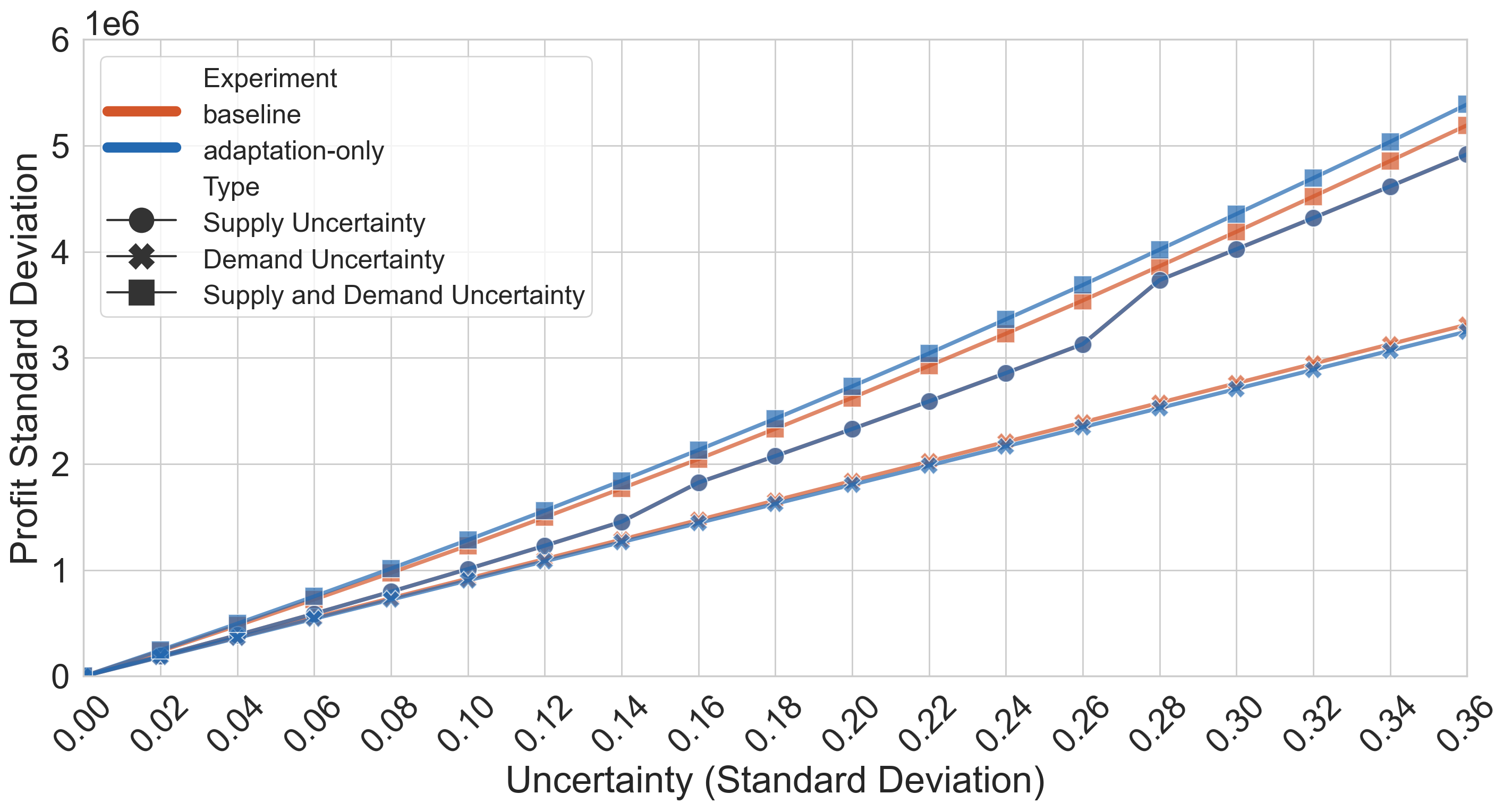}
    \caption{\centering Standard Deviation of Profit with Adaptation}
    \label{fig:adaptation-only_std}
\end{figure}

As seen earlier, adaptation can provide benefits to resilience by allowing multiple types of demand to be satisfied by a single good. To understand the benefits of adaptation, we model it as it is practiced in the industry, in the form of disabling cores and selling as a lower-end product. Because fusing (to disable cores) can be done any time after the part is packaged, we assume that the level of demand is known at the time adaptation is carried out.

Figure \ref{fig:adaptation-only_profit} shows the benefits of adaptation over the baseline. Unlike composition, there are no inherent benefits to profit with adaptation, so the profit when there is no variance is the same as the baseline case. For adaptation, we see $\lambda$ is 22.2\% when there is demand uncertainty, 8.6\% when there is supply and demand uncertainty, and 0.0\% when there is only supply uncertainty. 
This shows that while \textbf{adaptation provides a considerable benefit when only demand uncertainty is present}, there is disproportionately less benefit when both supply and demand uncertainties are present. This is because adaptation is limited in its ability to redistribute produced goods to demanded goods when there is limited supply (eg. a supply of 4-core can only satisfy 4-core demand, and a demand of 16-core can only be satisfied by a 16-core supply). {\bf There is no benefit from adaptation when there is only supply uncertainty.}

Results for standard deviation of profit are presented in Figure \ref{fig:adaptation-only_std}. Adaptation largely preserves the baseline standard deviation when there is demand uncertainty while providing significantly improved profits, signaling an improvement in variance of profit. There is negligible improvement in variance under the other two conditions.

\subsubsection{Fixing Composition’s problems through just-in-time packaging}

\begin{figure}[ht]
    \centering
    \captionsetup{skip=0pt}
    \includegraphics[width=1\linewidth]{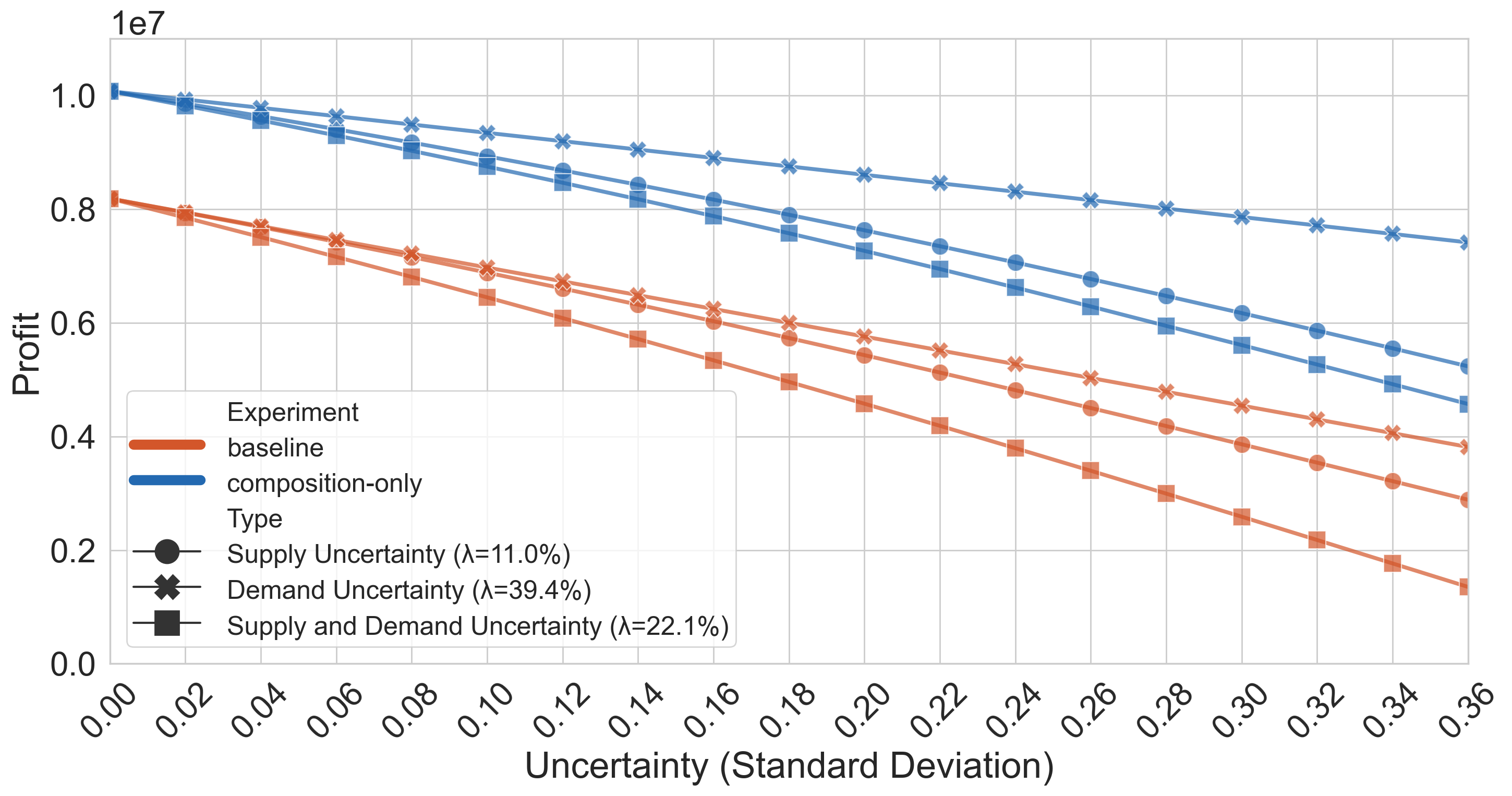}
    \caption{\centering Mean Profit with Just-in-time Packaging}
    \label{fig:stage1_composition-only_profit}
\end{figure}

\begin{figure}[ht]
    \centering
    \captionsetup{skip=0pt}
    \includegraphics[width=1\linewidth]{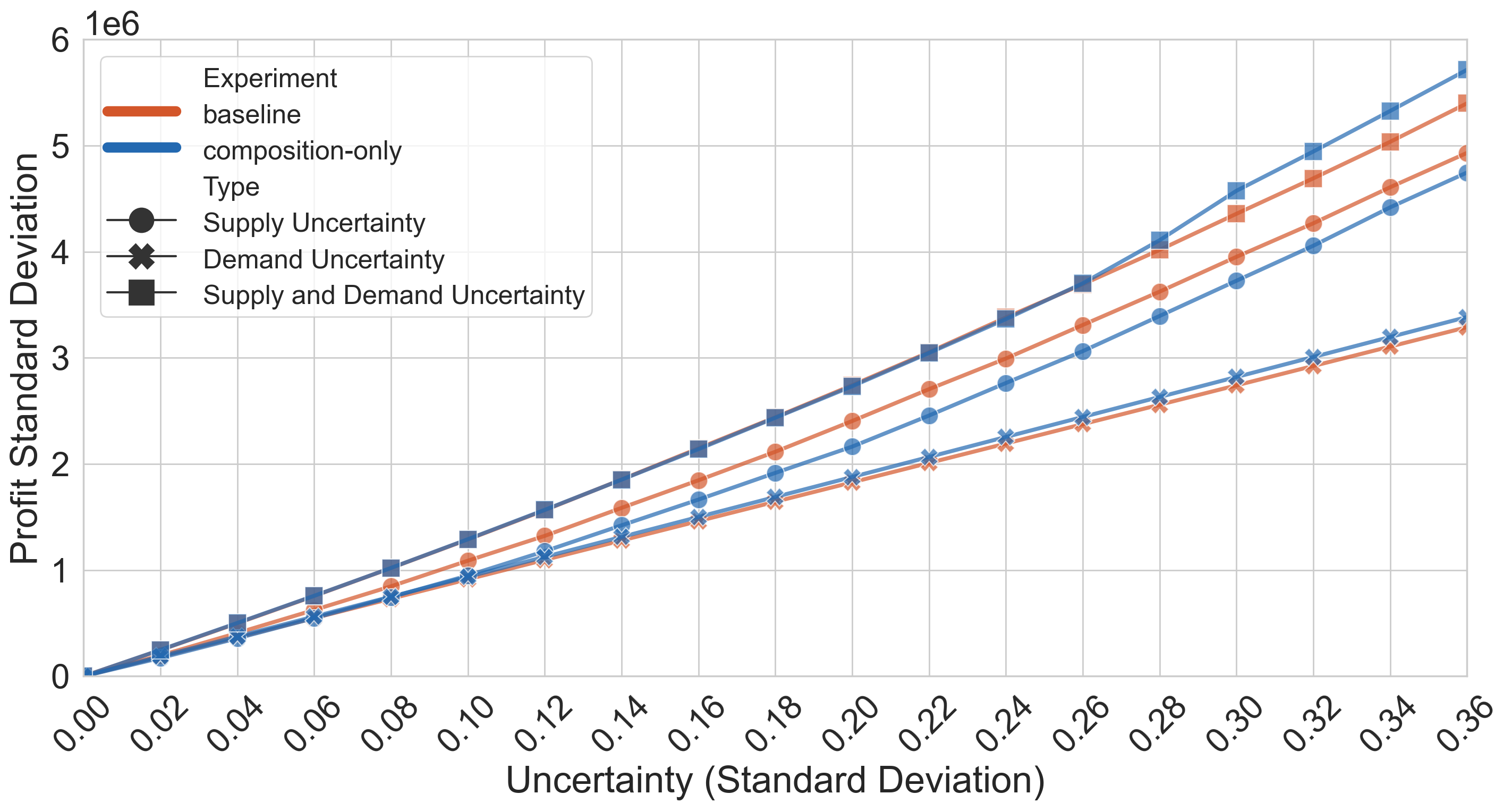}
    \caption{\centering Standard Deviation of Profit with Just-in-time Packaging}
    \label{fig:stage1_composition-only_std}
\end{figure}

In the previous analysis, it is seen that composition does not increase mean profit when demand uncertainty is present because it is unable to redistribute the received chiplets to meet variance in demand. This is because demand is unknown when the composition mapping takes place. So we ask, what may change if the level of demand is known at the time of packaging, perhaps through a breakthrough in distributed packaging at last-mile locations? This analysis is run the same as the composition analysis, except now the model knows demand while choosing the composition mappings.

Figure \ref{fig:stage1_composition-only_profit} shows results for mean profit under composition with just-in-time packaging. When factoring out the yield savings, we calculate a lambda of 39.4\% with demand uncertainty, 11.0\% with supply uncertainty and 22.1\% with both supply and demand uncertainty, showing \textbf{just-in-time composition significantly mitigates losses}. This is a significant improvement over baseline composition where there was virtually no benefit when demand uncertainty was present. 

 Figure \ref{fig:stage1_composition-only_std} shows standard deviation of profit with just-in-time composition. Profit-normalized variance is reduced in all cases, albeit to different degrees. This configuration maintains similar variance under demand and supply+demand uncertainty while increasing profit, and has a lower variance under supply uncertainty while providing greater profits.


\subsubsection{Enhancing Adaptation with Composition}
\begin{figure}[ht]
    \centering
    \captionsetup{skip=0pt}
    \includegraphics[width=1\linewidth]{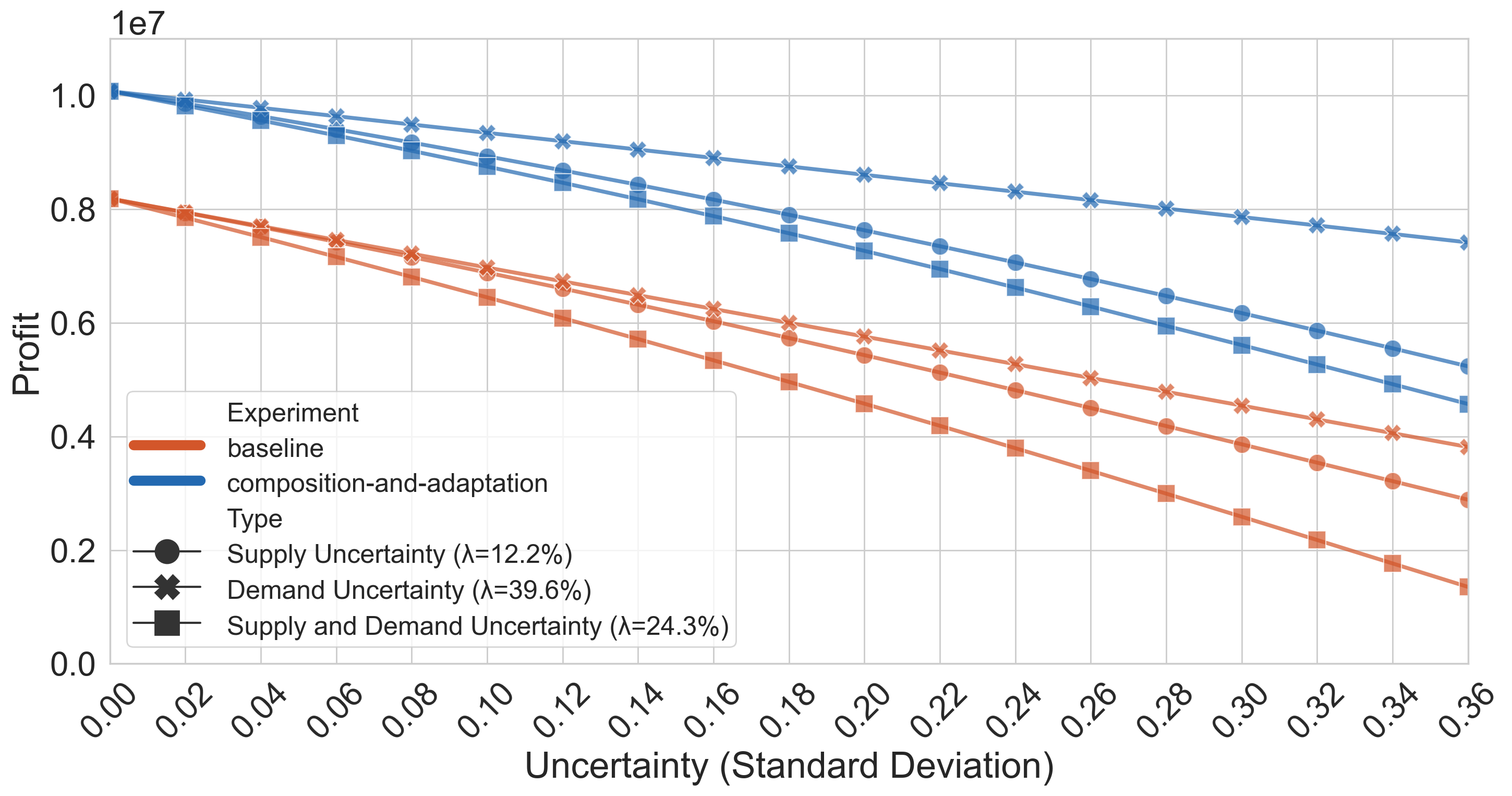}
    \caption{\centering Mean Profit with Adaptation + Composition}
    \label{fig:stage1_composition-and-adaptation_profit}
\end{figure}

\begin{figure}[ht]
    \centering
    \captionsetup{skip=0pt}
    \includegraphics[width=1\linewidth]{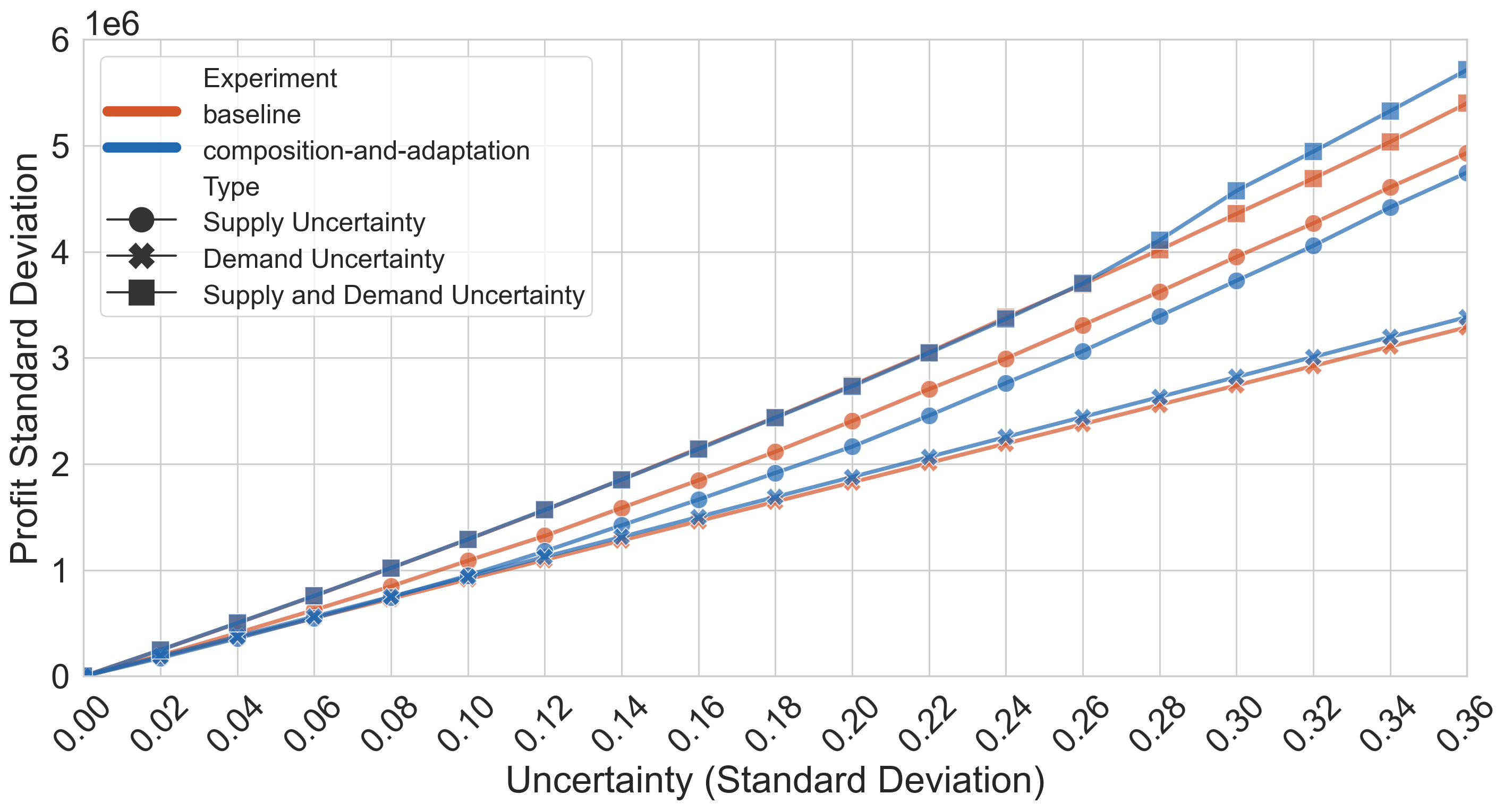}
    \caption{\centering Standard Deviation of Profit with Adaptation + Composition}
    \label{fig:stage1_composition-and-adaptation_std}
\end{figure}

Adaptation has limited efficacy when supply uncertainty is introduced (exclusively or in concert with demand uncertainty). But composition seems to provide some benefits here, so we explore if adding composition to adaptation may improve on baseline adaptation. We model composition with adaptation in the following way: It is the same as composition as described in the previous section, except after chips are packaged, it is still possible to disable cores and repurpose chips to meet multiple demands.

Figure \ref{fig:stage1_composition-and-adaptation_profit} shows mean profits using adaptation+composition. We see $\lambda$ is  39.6\% with demand uncertainty, 24.3\% with supply and demand uncertainty, and 12.2\% with supply uncertainty. These numbers represent an\textbf{ all-round improvement over baseline adaptation} and, in fact, even an improvement over just-in-time composition. Compared to just-in-time composition, $\lambda$ is 10 points higher in adaptation + composition for cases with only demand uncertainty and supply+demand uncertainty. Improvements to supply resilience are similar to the previous section, but there are improvements to demand resilience with this technique because the relation between the produced products and demanded products is fully connected. Eg. with baseline adaptation, a produced 8-core chip could be sold as either a 4 or 8-core chip. Now it can be composed to make a 16-core chip, sold as an 8-core, or adapted as a 4-core.  

Figure \ref{fig:stage1_composition-and-adaptation_std}  shows standard deviation of profit with adaptation and composition. This configuration maintains similar variance in profit under demand and supply+demand uncertainty while increasing profit, and lower variance in supply uncertainty while providing greater profits.

While adding composition to adaptation improves adaptation significantly, we continue to see minimal improvement to profit in the regime of supply uncertainty, and negligible reduction in profit variance. To address these drawbacks, we rely on {\em dispersion}.

\subsection{Dispersion}

Dispersion can improve resiliency by dispersing the production across suppliers who have a degree of independence with respect to the factors affecting their production. While dispersion is not strictly an architectural strategy, it can be enabled by architectural changes. For example, aggregating low-speed components in a design (eg. I/O) onto a separate chiplet will allow it to be manufactured on a trailing edge process (and therefore by more foundries), in effect 'dispersing' a portion of the overall design. 
There are two ways in which we model dispersion. In the first way, each of the produced chips is supplied from a different supplier. In the second way (dispersion2), we introduce two suppliers, and each produced chip is sourced from these two suppliers. We look at each of these two configurations separately. \\

\subsubsection{Using unique suppliers for each product}

\begin{figure}[ht]
    \centering
    \captionsetup{skip=0pt}
    \includegraphics[width=1\linewidth]{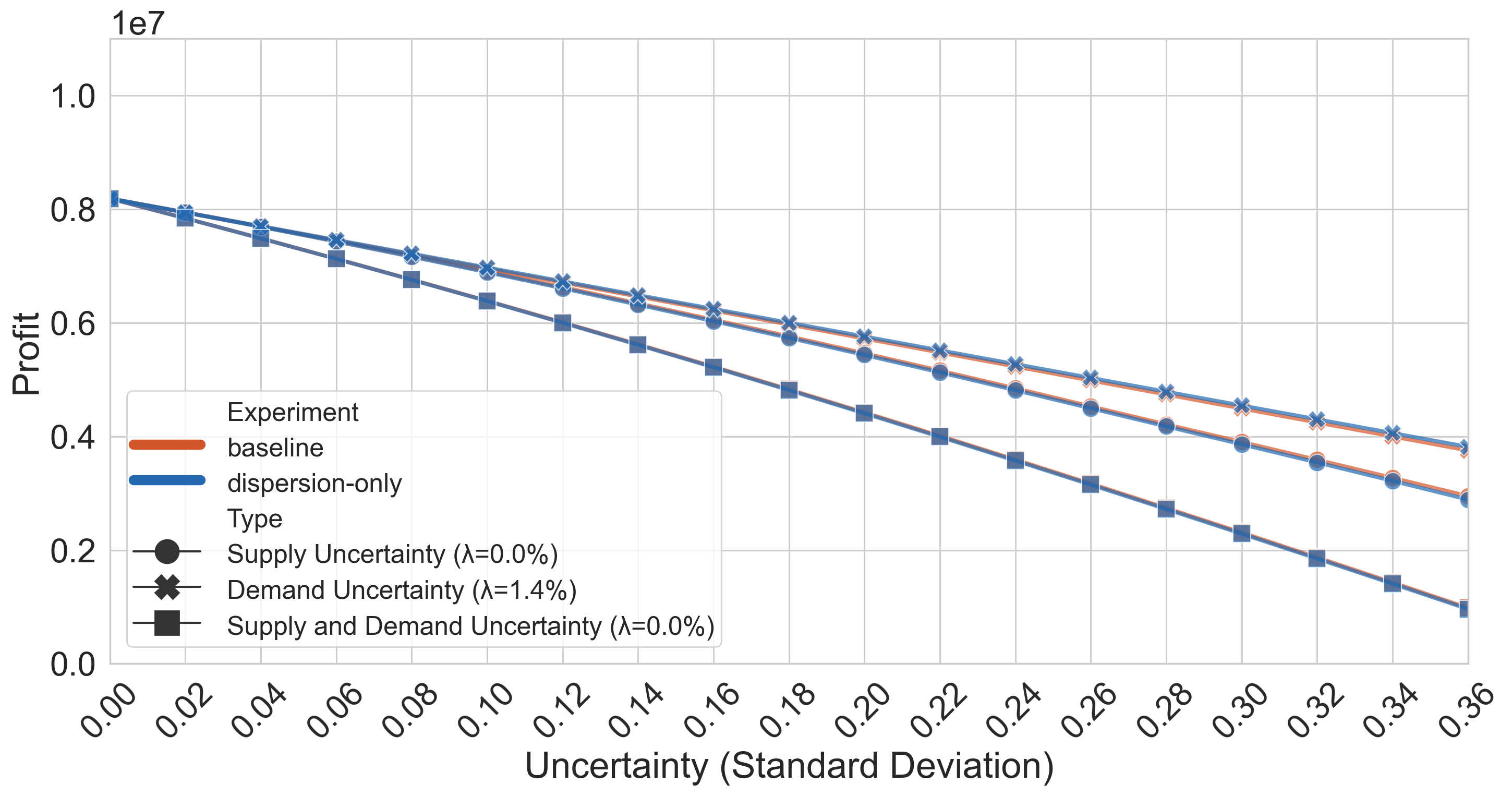}
    \caption{\centering Mean Profit with Dispersion}
    \label{fig:dispersion-only_profit}
\end{figure}

\begin{figure}[ht]
    \centering
    \captionsetup{skip=0pt}
    \includegraphics[width=1\linewidth]{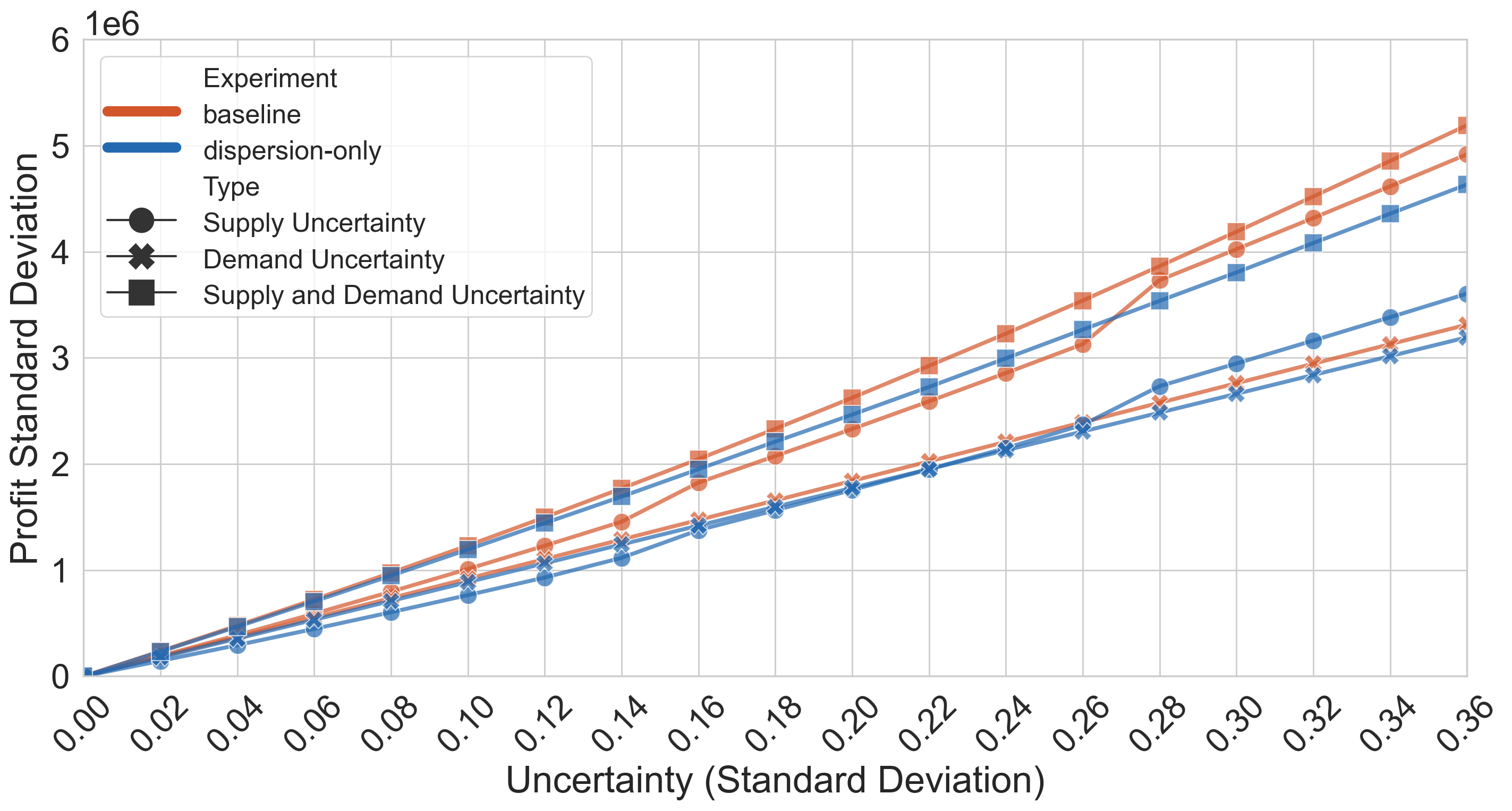}
    \caption{\centering Standard Deviation of Profit with Dispersion}
    \label{fig:dispersion-only_std}
\end{figure}

\begin{figure}[ht]
    \centering
    \captionsetup{skip=0pt}
    \includegraphics[width=1\linewidth]{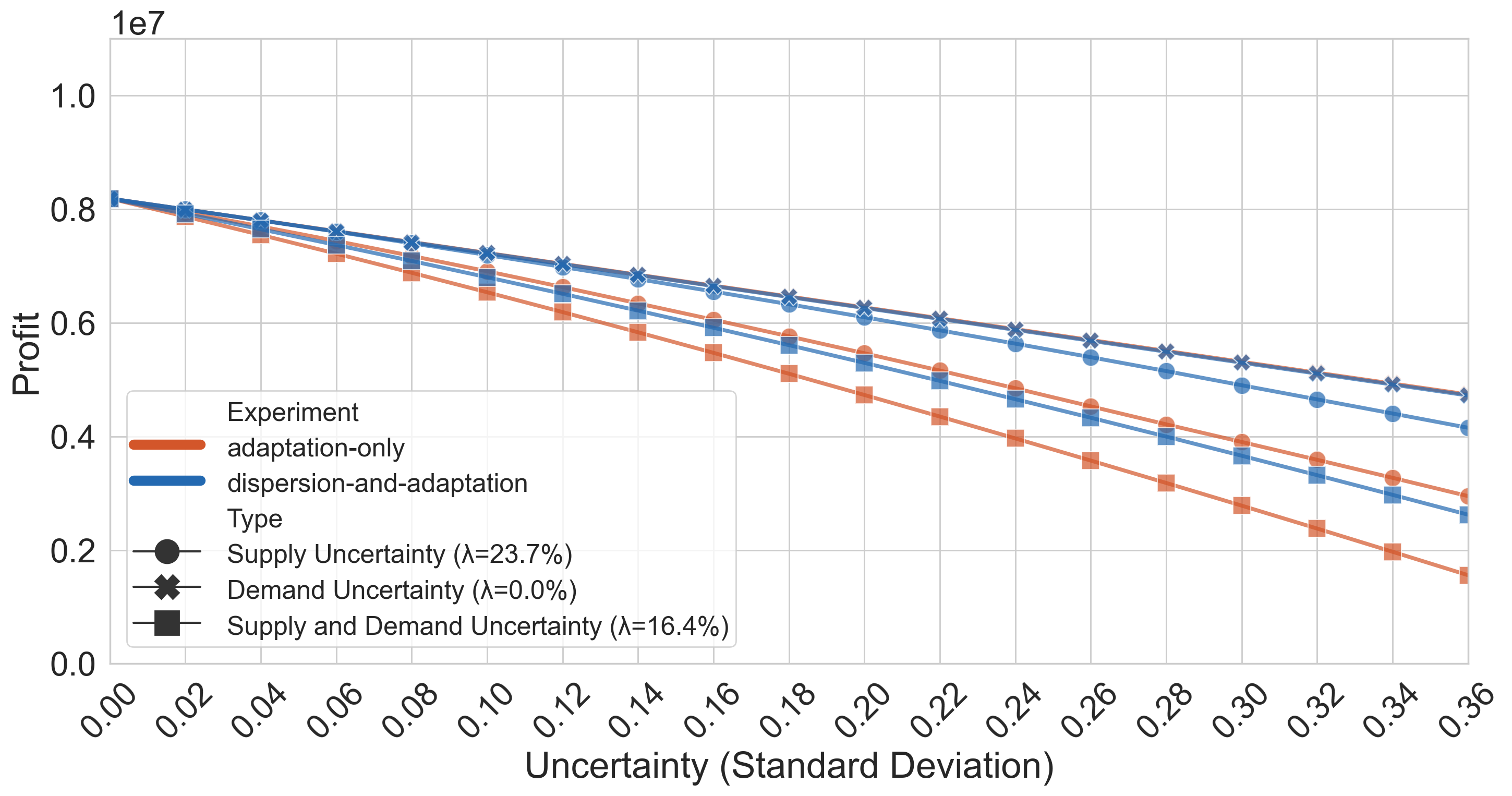}
    \caption{\centering Mean Profit with Dispersion + Adaptation}
    \label{fig:dispersion-and-adaptation_profit}
\end{figure}

\begin{figure}[ht]
    \centering
    \captionsetup{skip=0pt}
    \includegraphics[width=1\linewidth]{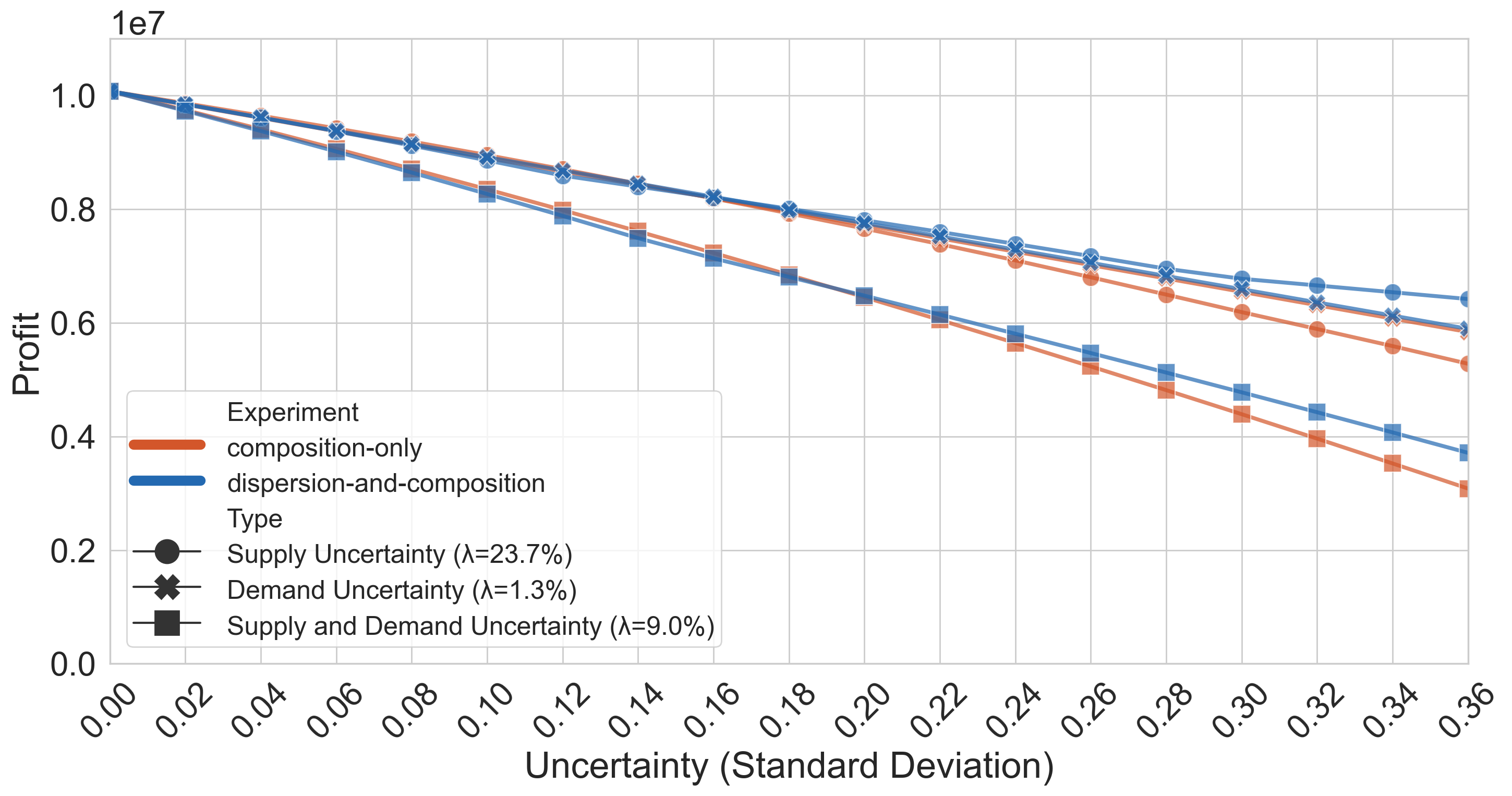}
    \caption{\centering Mean Profit with Dispersion + Composition}
    \label{fig:dispersion-and-composition_profit}
\end{figure}

Using unique suppliers for each product decouples the supplies of each of the products from each other, potentially offering improved resiliency against swings in supply. Concretely, this means three different supply distributions are used to sample each value in $\zs$ corresponding to each produced chip. 

Profit results for dispersion alone are presented in Figure \ref{fig:dispersion-only_profit}. We notice \textbf{this provides no benefit over the baseline for profit}. This is because surpluses in the supply of one product cannot make up for deficits in the supply of the other since chips cannot be transformed in the baseline case. However, when transformations between produced and demanded goods are enabled by adding dispersion to adaptation (Figure \ref{fig:dispersion-and-adaptation_profit} ) or to composition (Figure \ref{fig:dispersion-and-composition_profit} ) we see significant results. We compare these to baseline composition and adaptation. Adding dispersion to composition produces a $\lambda$ of 23.7\% and 9.0\% in supply and supply + demand uncertain markets, respectively. For adaptation with dispersion, these values are 23.7\% and 16.4\% respectively.

Even though there were no improvements to profit from dispersion only, there was a \textbf{36.9\% reduction in profit standard deviation under supply uncertainty} and a 20.7\% reduction in profit standard deviation under supply + demand uncertainty (Figure \ref{fig:dispersion-only_std}). A similar improvement is seen for dispersion profit standard deviation when used with composition and adaptation.

\subsubsection{Using two suppliers across each product}

\begin{figure}[ht]
    \centering
    \captionsetup{skip=0pt}
    \includegraphics[width=1\linewidth]{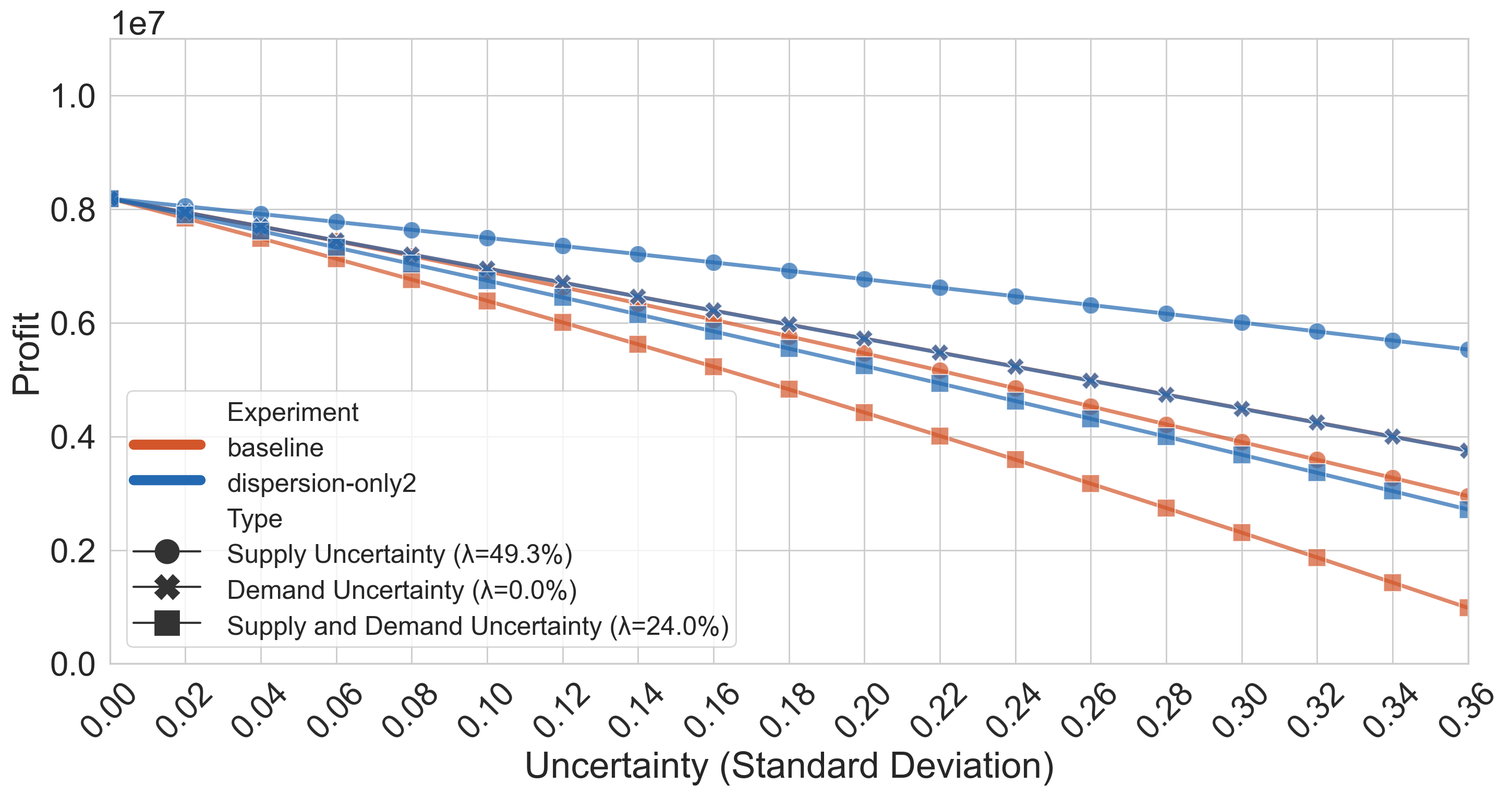}
    \caption{\centering Mean Profit with Dispersion2}
    \label{fig:dispersion-only_profit2}
\end{figure}

\begin{figure}[ht]
    \centering
    \captionsetup{skip=0pt}
    \includegraphics[width=1\linewidth]{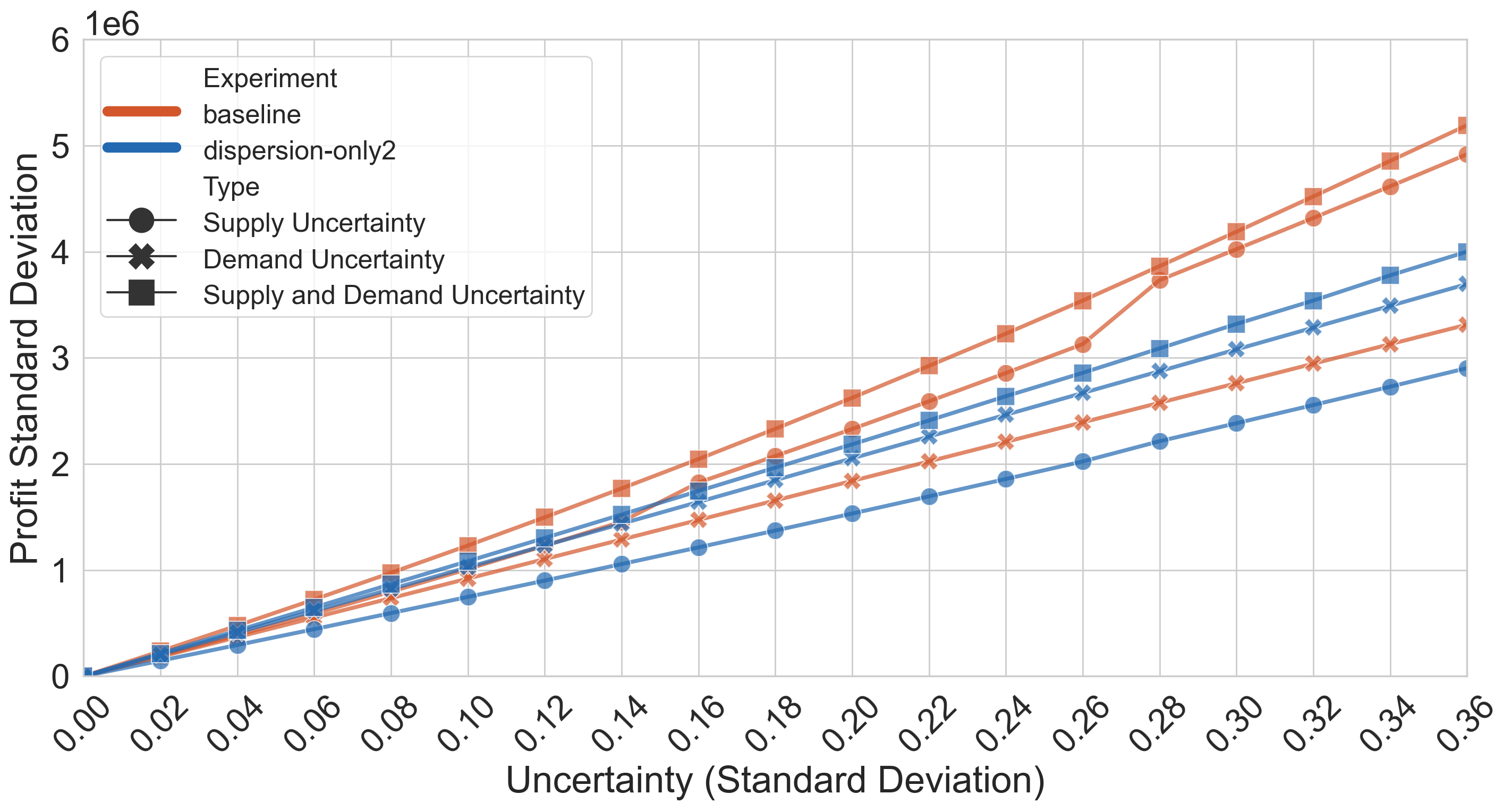}
    \caption{\centering Standard Deviation of Profit with Dispersion2}
    \label{fig:dispersion-only2_std}
\end{figure}

\begin{figure}[ht]
    \centering
    \captionsetup{skip=0pt}
    \includegraphics[width=1\linewidth]{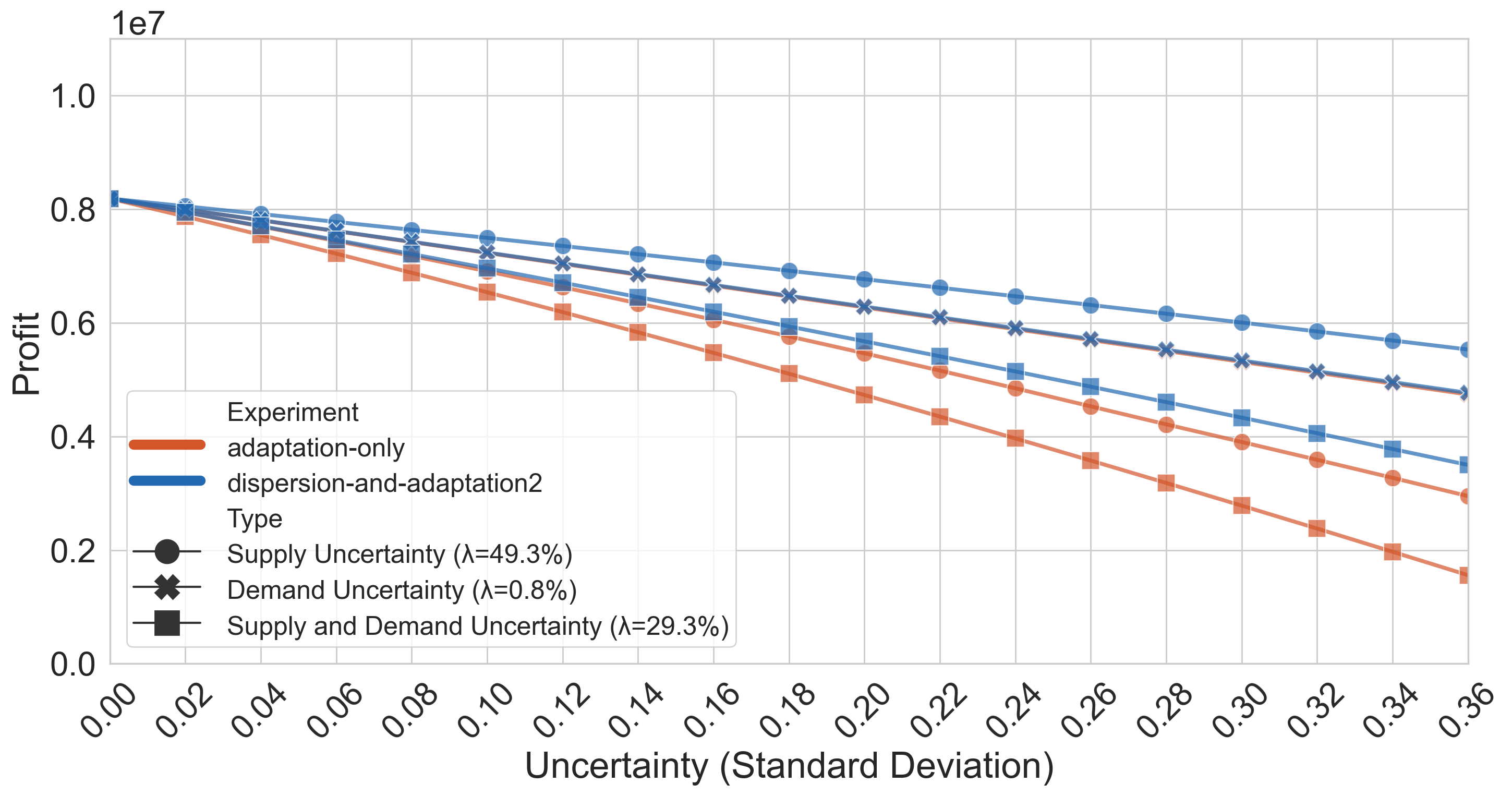}
    \caption{\centering Mean Profit with Dispersion2 + Adaptation}
    \label{fig:dispersion-and-adaptation2_profit}
\end{figure}

\begin{figure}[ht]
    \centering
    \captionsetup{skip=0pt}
    \includegraphics[width=1\linewidth]{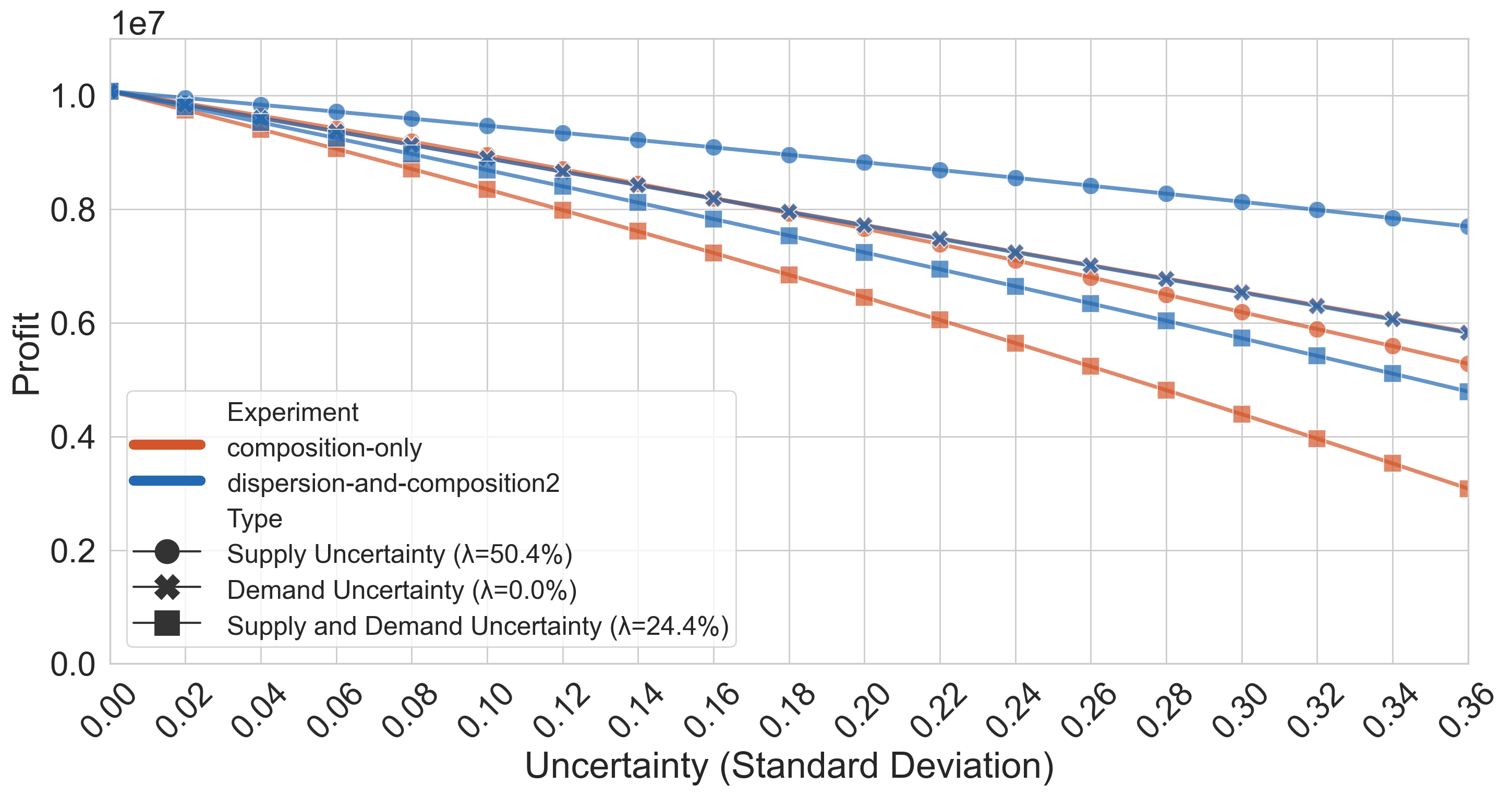}
    \caption{\centering  Mean Profit with Dispersion2 + Composition}
    \label{fig:dispersion-and-composition2_profit}
\end{figure}

Under this strategy, there are two independent suppliers and all products in the market can be sourced from either one of them. Concretely, this translates to two independent supply distributions each operating on half the goods ordered. The two suppliers are equal in all ways, so we force the model to order an equal quantity from both of them. 

Baseline results depicted in Figure \ref{fig:dispersion-only_profit2} show that a multi-sourcing strategy has benefits even without other architectural interventions. \textbf{Results show an impressive 49.3\% $\lambda$ in supply-uncertain markets and a 24.0\% $\lambda$ for supply + demand.} Significant improvements are seen in profit variance as well. 
Multi-sourcing dispersion used with composition (Figure  \ref{fig:dispersion-and-composition2_profit}) or adaptation (Figure \ref{fig:dispersion-and-adaptation2_profit}) continues to provide improvements over their respective baselines. For dispersion + adaptation we see a $\lambda$ of 49.3\% and 29.3\% under supply and supply+demand uncertainty, respectively, and for dispersion +composition these numbers are 50.4\% and 24.4\%, respectively. These improvements are much greater than that seen with the previous dispersion technique, yet similar to the baseline case of multi-supplier dispersion, suggesting that \textbf{composition and adaptation do not further enhance multi-sourcing dispersion under supply uncertainty}. 

For standard deviation in profit, we see a reduction between 25\% to 40.8\%  under the multi-sourcing strategy in the baseline (Figure \ref{fig:dispersion-only2_std}), and these improvements largely stay the same when composition and adaptation are added. 

\begin{figure}[ht]
    \centering
    \captionsetup{skip=0pt}
    \includegraphics[width=1\linewidth]{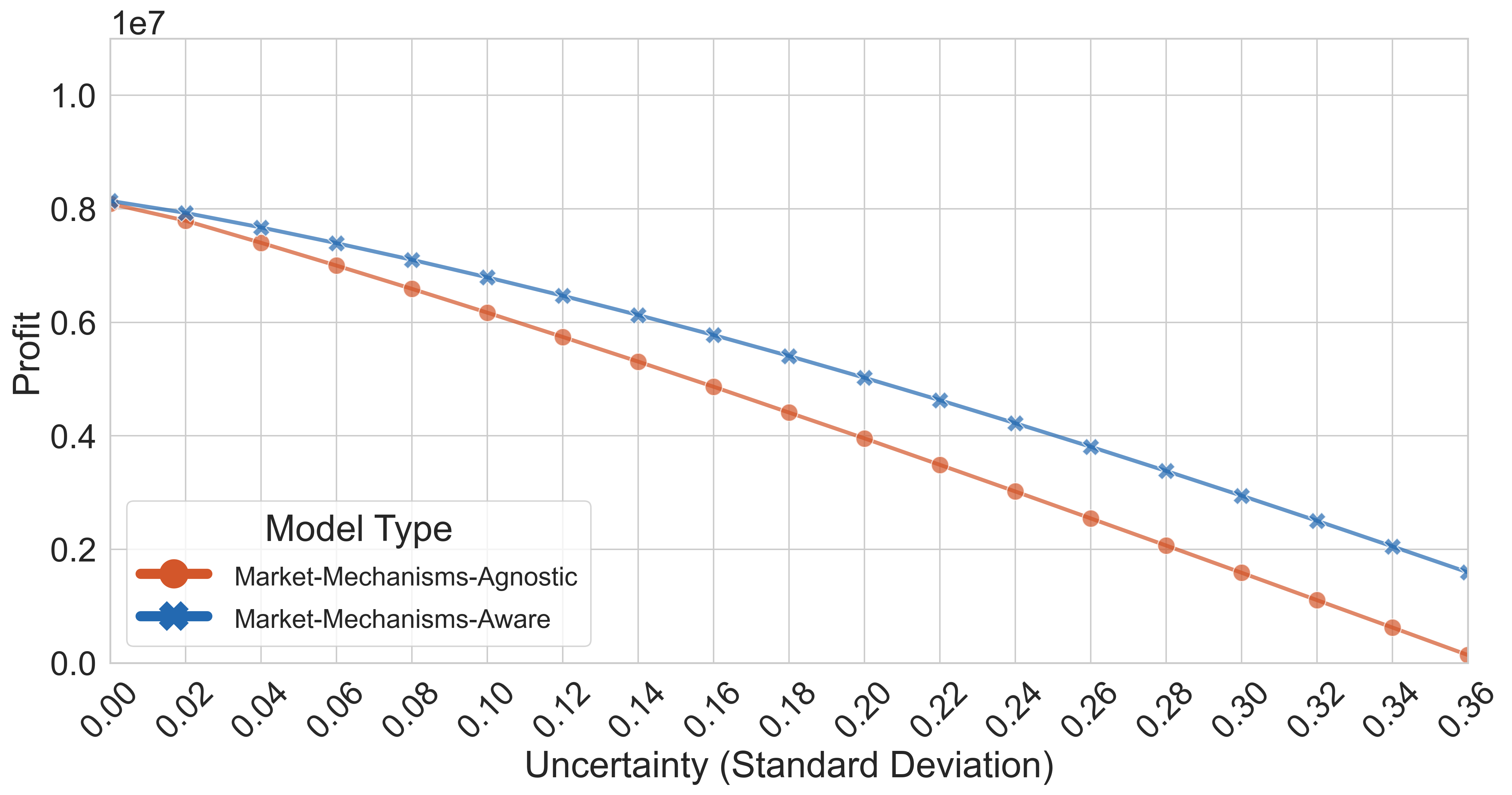}
    \caption{\centering Effect of Knowledge of Market Mechanism on Mean Profit}
    \label{fig::Model-0-vs-2_const-shortage}
\end{figure}

\subsection{Market Mechanisms}

Till now, our experiments have modeled a market for semiconductor chips statically without any consideration for market mechanisms. But what happens when market mechanisms like competition and/or elasticity of demand are considered? Would the efficacy of any of the techniques we proposed previously change? To explore this, we introduce a partial equilibrium-based~\cite{gilbert_2017} model of market dynamics to our model; partial equilibrium can model the first-order effects of competition, production and demand. This is incorporated into our model by integrating a demand curve. With this addition, the unit benefit that a good is sold at is linearly related to number of goods purchased. The new unit benefit is given by:

\begin{align*}
    \text{Unit Benefit}\ =&\ \text{(Goods Sold - Base Demand)}*\\
    &\ \text{Market Elasticity}+ \text{Base Price}
\end{align*}

where base demand is the same as it was in the baseline case, and base price is the same as unit benefit in the previous model. Market elasticity signifies how elastic the demand is (i.e the extent to which demand changes with a change in price). To ensure comparability between the two models, the parameter values for elasticity for each of the different chips were chosen such that the optimal point in the demand curve is at the point where base demand number of chips are sold, each at base price, yielding the same revenue as in the previous model. Combining HP's PC sales quantities from figure \ref{fig:PC_mar_fluct} and the corresponding revenue data from HP \cite{hp-rev-q1-2018, hp-rev-q2-2018, hp-rev-q3-2018, hp-rev-q4-2018, hp-rev-q1-2019, hp-rev-q2-2019, hp-rev-q3-2019, hp-rev-q4-2019, hp-rev-q1-2020, hp-rev-q2-2020, hp-rev-q3-2020, hp-rev-q4-2020, hp-rev-q1-2021, hp-rev-q2-2021, hp-rev-q3-2021, hp-rev-q4-2021, hp-rev-q1-2022, hp-rev-q2-2022}, and assuming chips comprise 10\% of the cost of a PC, we see a similar magnitude of elasticity to our calculated parameter values. 
Demand uncertainty is modeled through shifting the demand curve left or right by scaling base demand by the demand factor. The supply side of the model remains unchanged. To ensure consistency between the models, shortage cost is also modeled in the same way. The model is aware of the market mechanism and optimizes for highest expected profit accordingly. 

Results on the effect of considering market mechanisms are presented in Figure \ref{fig::Model-0-vs-2_const-shortage}. Results are presented for supply + demand uncertainty case. Despite the differing shapes (linear - market mechanism-agnostic vs negative exponential - market-mechanism aware), the two curves are relatively close to each other and come closer as uncertainty increases. The calculated $\lambda$ value (when considering knowledge of uncertainty as an intervention) is 30\% at low values of uncertainty, and this reduces to 18.8\% at the high end of uncertainty variance. This suggests that \textbf{knowledge of market mechanisms is largely equivalent to adding a low-performing intervention in terms of profit benefits}. As we have shown, interventions can be composed to improve results, and the reason knowledge of market mechanisms is making a loss is 
the same 
as in the baseline model, except losses are padded by increased selling prices (due to awareness of the market)  when supply is low or demand is high. Given this, we conclude that all our previous conclusions continue to be interesting and valid when market mechanisms are considered. \\

\section{Summary and Conclusions}
We presented the first study on the role of chip architecture in improving semiconductor supply chain resilience. We first found that one-to-one chip-to-demand mapping, as is common today, is a significant source of fragility, leading to over a three-fourth reduction in profits at our estimates of industry supply and demand uncertainty. As currently done, the architectural strategy of {\em composition} is ineffective in improving profits beyond the yield savings it offers, but \emph{just-in-time composition} mitigates up to 33\% of the losses from demand volatility. This also mitigates some supply volatility by directing (limited) supply to the highest margin product.  The strategy of {\em adaptation} can mitigate 20\% of losses from demand uncertainty but is unable to mitigate losses from supply volatility. Moreover, adding composition on top of adaptation can double its effectiveness in demand-volatile situations and provide some benefits in supply-volatile situations. In developing two flavors of dispersion, we showed that multi-sourcing dispersion is superior, mitigating nearly half of the losses from supply uncertainty.  Finally, we showed that knowledge of how the market behaves helps improve profits but less than most of our proposed interventions, and in fact these interventions continue to provide benefits when market mechanisms are considered. Overall, nearly half of the losses from both supply and demand volatility can be mitigated using the techniques we propose. 

\bibliographystyle{IEEEtranS}
\bibliography{refs}

\begin{thebibliography}{10}
\providecommand{\url}[1]{#1}
\csname url@samestyle\endcsname
\providecommand{\newblock}{\relax}
\providecommand{\bibinfo}[2]{#2}
\providecommand{\BIBentrySTDinterwordspacing}{\spaceskip=0pt\relax}
\providecommand{\BIBentryALTinterwordstretchfactor}{4}
\providecommand{\BIBentryALTinterwordspacing}{\spaceskip=\fontdimen2\font plus
\BIBentryALTinterwordstretchfactor\fontdimen3\font minus
  \fontdimen4\font\relax}
\providecommand{\BIBforeignlanguage}[2]{{%
\expandafter\ifx\csname l@#1\endcsname\relax
\typeout{** WARNING: IEEEtranS.bst: No hyphenation pattern has been}%
\typeout{** loaded for the language `#1'. Using the pattern for}%
\typeout{** the default language instead.}%
\else
\language=\csname l@#1\endcsname
\fi
#2}}
\providecommand{\BIBdecl}{\relax}
\BIBdecl

\bibitem{IDC_Semi_Rev_Forc_19}
\BIBentryALTinterwordspacing
``After three consecutive years of growth, idc forecasts worldwide
  semiconductor revenue to decline by 7.2\% in 2019.'' [Online]. Available:
  \url{https://web.archive.org/web/20221122031546/https://www.bloomberg.com/press-releases/2019-05-15/after-three-consecutive-years-of-growth-idc-forecasts-worldwide-semiconductor-revenue-to-decline-by-7-2-in-2019}
\BIBentrySTDinterwordspacing

\bibitem{4core2}
\BIBentryALTinterwordspacing
``Amd ryzen 3 3300x benchmark, test and specs.'' [Online]. Available:
  \url{https://www.cpu-monkey.com/en/cpu-amd_ryzen_3_3300x}
\BIBentrySTDinterwordspacing

\bibitem{6core1}
\BIBentryALTinterwordspacing
``Amd ryzen 5 3600 vs amd ryzen 5 3600x benchmark, comparison and
  differences.'' [Online]. Available:
  \url{https://www.cpu-monkey.com/en/compare_cpu-amd_ryzen_5_3600-vs-amd_ryzen_5_3600x}
\BIBentrySTDinterwordspacing

\bibitem{t9}
\BIBentryALTinterwordspacing
``Asahi kasei semiconductor factory fire resumption time time substitute
  production request to other companies.'' [Online]. Available:
  \url{https://www.tellerreport.com/business/2021-02-09-\%0A---asahi-kasei-semiconductor-factory-fire-resumption-time-time-substitute-production-request-to-other-companies-\%0A--.Hk8_mw31Wu.html}
\BIBentrySTDinterwordspacing

\bibitem{semi_utilization}
\BIBentryALTinterwordspacing
``Capacity utilization: Manufacturing: Durable goods: Semiconductor and other
  electronic component (naics = 3344).'' [Online]. Available:
  \url{https://www.federalreserve.gov/releases/g17/Revisions/20220628/ipdisk/alltables.txt}
\BIBentrySTDinterwordspacing

\bibitem{noauthor_coronavirus_nodate}
\BIBentryALTinterwordspacing
``Coronavirus: {Implications} for semiconductor demand {\textbar} {McKinsey}.''
  [Online]. Available:
  \url{https://www.mckinsey.com/industries/semiconductors/our-insights/coronavirus-implications-for-the-semiconductor-industry}
\BIBentrySTDinterwordspacing

\bibitem{noauthor_gartner_nodate}
\BIBentryALTinterwordspacing
``\BIBforeignlanguage{en}{Gartner {Says} {Global} {Smartphone} {Sales}
  {Declined} 20\% in {Second} {Quarter} of 2020}.'' [Online]. Available:
  \url{https://www.gartner.com/en/newsroom/press-releases/2020-08-25-gartner-says-global-smartphone-sales-declined-20--in-}
\BIBentrySTDinterwordspacing

\bibitem{PC_shipments_1Q20}
\BIBentryALTinterwordspacing
``Gartner says worldwide pc shipments declined 12.3 percent in the first
  quarter of 2020 due to coronavirus pandemic,'' \emph{Gartner}. [Online].
  Available:
  \url{https://www.gartner.com/en/newsroom/press-releases/2020-04-13-gartner-says-worldwide-pc-shipments-declined-12-point-3-percent-in-the-first-quarter-of-2020-due-to-coronavirus-pandemic}
\BIBentrySTDinterwordspacing

\bibitem{PC_shipments_1Q19}
\BIBentryALTinterwordspacing
``Gartner says worldwide pc shipments declined 4.6 percent,'' \emph{Gartner}.
  [Online]. Available:
  \url{https://www.gartner.com/en/newsroom/press-releases/2019-04-10-gartner-says-worldwide-pc-shipments-declined-4-6-perc}
\BIBentrySTDinterwordspacing

\bibitem{PC_shipments_4Q21}
\BIBentryALTinterwordspacing
``Gartner says worldwide pc shipments declined 5 percent in fourth quarter of
  2021 but grew nearly 10 percent for the year,'' \emph{Gartner}. [Online].
  Available:
  \url{https://www.gartner.com/en/newsroom/press-releases/2022-01-12-gartner-says-worldwide-pc-shipments-declined-5-percent-in-fourth-quarter-of-2021-but-grew-nearly-10-percent-for-the-year}
\BIBentrySTDinterwordspacing

\bibitem{PC_shipments_1Q22}
\BIBentryALTinterwordspacing
``Gartner says worldwide pc shipments declined 7 percent in first quarter of
  2022,'' \emph{Gartner}. [Online]. Available:
  \url{https://www.gartner.com/en/newsroom/press-releases/2022-04-11-gartner-says-worldwide-pc-shipments-declined-7-percent-in-first-quarter-of-2022}
\BIBentrySTDinterwordspacing

\bibitem{PC_shipments_2Q22}
\BIBentryALTinterwordspacing
``Gartner says worldwide pc shipments experienced the sharpest decline in nine
  years in second quarter of 2022,'' \emph{Gartner}. [Online]. Available:
  \url{https://www.gartner.com/en/newsroom/press-releases/2022-07-11-gartner-says-worldwide-pc-shipments-experienced-the-sharpest-decline-in-nine-years-in-second-quarter-of-2022}
\BIBentrySTDinterwordspacing

\bibitem{PC_shipments_3Q21}
\BIBentryALTinterwordspacing
``Gartner says worldwide pc shipments grew 1 percent in third quarter of
  2021,'' \emph{Gartner}. [Online]. Available:
  \url{https://www.gartner.com/en/newsroom/press-releases/2021-10-11-gartner-says-worldwide-pc-shipments-grew-1-percent-in-third-quarter-of-2021}
\BIBentrySTDinterwordspacing

\bibitem{PC_shipments_4Q20}
\BIBentryALTinterwordspacing
``Gartner says worldwide pc shipments grew 10.7 percent in the fourth quarter
  of 2020 and 4 point 8 percent for the year,'' \emph{Gartner}. [Online].
  Available:
  \url{https://www.gartner.com/en/newsroom/press-releases/2021-01-11-gartner-says-worldwide-pc-shipments-grew-10-point-7-percent-in-the-fourth-quarter-of-2020-and-4-point-8-percent-for-the-year}
\BIBentrySTDinterwordspacing

\bibitem{PC_shipments_3Q19}
\BIBentryALTinterwordspacing
``Gartner says worldwide pc shipments grew 1.1 percent in third quarter of
  2019,'' \emph{Gartner}. [Online]. Available:
  \url{https://www.gartner.com/en/newsroom/press-releases/2019-10-10-gartner-says-worldwide-pc-shipments-grew-1point1-percent-in-third-quarter-of-2019}
\BIBentrySTDinterwordspacing

\bibitem{PC_shipments_2Q19}
\BIBentryALTinterwordspacing
``Gartner says worldwide pc shipments grew 1.5 percent in second quarter of
  2019,'' \emph{Gartner}. [Online]. Available:
  \url{https://www.gartner.com/en/newsroom/press-releases/2019-07-11-gartner-says-worldwide-pc-shipments-grew-1point5percent-in-second-quarter-of-2019}
\BIBentrySTDinterwordspacing

\bibitem{PC_shipments_4Q19}
\BIBentryALTinterwordspacing
``Gartner says worldwide pc shipments grew 2.3 percent in 4q19 and .6 percent
  for the year,'' \emph{Gartner}. [Online]. Available:
  \url{https://www.gartner.com/en/newsroom/press-releases/2020-01-13-gartner-says-worldwide-pc-shipments-grew-2-point-3-percent-in-4q19-and-point-6-percent-for-the-year}
\BIBentrySTDinterwordspacing

\bibitem{PC_shipments_2Q20}
\BIBentryALTinterwordspacing
``Gartner says worldwide pc shipments grew 2.8 percent in second quarter of
  2020,'' \emph{Gartner}. [Online]. Available:
  \url{https://www.gartner.com/en/newsroom/press-releases/2020-07-09-gartner-says-worldwide-pc-shipments-grew-2point8-percent-in-second-quarter-of-2020}
\BIBentrySTDinterwordspacing

\bibitem{PC_shipments_1Q21}
\BIBentryALTinterwordspacing
``Gartner says worldwide pc shipments grew 32 percent in first quarter of
  2021,'' \emph{Gartner}. [Online]. Available:
  \url{https://www.gartner.com/en/newsroom/press-releases/2021-04-12-gartner-says-worldwide-pc-shipments-grew-32-percent-in-first-quarter-of-2021}
\BIBentrySTDinterwordspacing

\bibitem{PC_shipments_3Q20}
\BIBentryALTinterwordspacing
``Gartner says worldwide pc shipments grew 3.6 percent in the third quarter of
  2020,'' \emph{Gartner}. [Online]. Available:
  \url{https://www.gartner.com/en/newsroom/press-releases/2020-10-12-gartner-says-worldwide-pc-shipments-grew-3-point-six-percent-in-the-third-quarter-of-2020}
\BIBentrySTDinterwordspacing

\bibitem{PC_shipments_2Q21}
\BIBentryALTinterwordspacing
``Gartner says worldwide pc shipments grew 4 point six in second quarter of
  2021,'' \emph{Gartner}. [Online]. Available:
  \url{https://www.gartner.com/en/newsroom/press-releases/2021-07-12-gartner-says-worldwide-pc-shipments-grew-4-point-six-in-second-quarter-of-2021}
\BIBentrySTDinterwordspacing

\bibitem{semi_revenue}
\BIBentryALTinterwordspacing
``Historical billings report.'' [Online]. Available:
  \url{https://www.wsts.org/esraCMS/extension/media/f/WST/5711/WSTS-Historical-Billings-Report-Sep2022data.xlsx}
\BIBentrySTDinterwordspacing

\bibitem{8core}
\BIBentryALTinterwordspacing
``Intel core i7-7700k vs amd ryzen 3 3100 benchmark, comparison and
  differences.'' [Online]. Available:
  \url{https://www.cpu-monkey.com/en/compare_cpu-intel_core_i7_7700k-vs-amd_ryzen_3_3100}
\BIBentrySTDinterwordspacing

\bibitem{6core2}
\BIBentryALTinterwordspacing
``Intel core i7-8700k benchmark, test and specs.'' [Online]. Available:
  \url{https://www.cpu-monkey.com/en/cpu-intel_core_i7_8700k}
\BIBentrySTDinterwordspacing

\bibitem{4core1}
\BIBentryALTinterwordspacing
``Intel core i9-9900k vs intel core i7-10700k benchmark, comparison and
  differences.'' [Online]. Available:
  \url{https://www.cpu-monkey.com/en/compare_cpu-intel_core_i9_9900k-vs-intel_core_i7_10700k}
\BIBentrySTDinterwordspacing

\bibitem{IDC_Semi_Rev_Forc_18}
\BIBentryALTinterwordspacing
``Market growth rate peaks after a strong 2017; idc forecasts semiconductor
  revenue growth of 7.7\%, reaching \$450 billion in 2018.'' [Online].
  Available:
  \url{https://web.archive.org/web/20221122031651/https://www.bloomberg.com/press-releases/2018-05-21/market-growth-rate-peaks-after-a-strong-2017-idc-forecasts-semiconductor-revenue-growth-of-7-7-reaching-450-billion-in-2018}
\BIBentrySTDinterwordspacing

\bibitem{IDC_Semi_Rev_Forc_20}
\BIBentryALTinterwordspacing
``Semiconductor downturn to continue into 2020 due to covid-19 impact; idc
  forecasts worldwide non-memory semiconductor revenue to decline 7.2\% in
  2020.'' [Online]. Available:
  \url{https://web.archive.org/web/20210912143858/https://www.businesswire.com/news/home/20200504005178/en/Semiconductor-Downturn-to-Continue-Into-2020-Due-to-COVID-19-Impact-IDC-Forecasts-Worldwide-Non-Memory-Semiconductor-Revenue-to-Decline-7.2-in-2020}
\BIBentrySTDinterwordspacing

\bibitem{Gartner_Semi_Rev_Forc_2Q18}
\BIBentryALTinterwordspacing
``Semiconductor forecast database, worldwide, 2q18 update,'' \emph{Gartner}.
  [Online]. Available: \url{https://www.gartner.com/en/documents/3881084}
\BIBentrySTDinterwordspacing

\bibitem{Gartner_Semi_Rev_Forc_2Q19}
\BIBentryALTinterwordspacing
``Semiconductor forecast database, worldwide, 2q19 update,'' \emph{Gartner}.
  [Online]. Available: \url{https://www.gartner.com/en/documents/3942113}
\BIBentrySTDinterwordspacing

\bibitem{Gartner_Semi_Rev_Forc_2Q20}
\BIBentryALTinterwordspacing
``Semiconductor forecast database, worldwide, 2q20 update,'' \emph{Gartner}.
  [Online]. Available: \url{https://www.gartner.com/en/documents/3986978}
\BIBentrySTDinterwordspacing

\bibitem{Gartner_Semi_Rev_Forc_2Q21}
\BIBentryALTinterwordspacing
``Semiconductor forecast database, worldwide, 2q21 update,'' \emph{Gartner}.
  [Online]. Available: \url{https://www.gartner.com/en/documents/4003111}
\BIBentrySTDinterwordspacing

\bibitem{hp-rev-q1-2018}
\BIBentryALTinterwordspacing
``Hp inc.. and subsidiaries consolidated condensed statements of earnings,''
  2018. [Online]. Available:
  \url{https://s2.q4cdn.com/602190090/files/doc_financials/2018/q1/q1-2018-quarterly-results.pdf}
\BIBentrySTDinterwordspacing

\bibitem{hp-rev-q2-2018}
\BIBentryALTinterwordspacing
``Hp inc.. and subsidiaries consolidated condensed statements of earnings,''
  2018. [Online]. Available:
  \url{https://s2.q4cdn.com/602190090/files/doc_financials/2018/q2/q2-2018-quarterly-results.pdf}
\BIBentrySTDinterwordspacing

\bibitem{hp-rev-q3-2018}
\BIBentryALTinterwordspacing
``Hp inc.. and subsidiaries consolidated condensed statements of earnings,''
  2018. [Online]. Available:
  \url{https://s2.q4cdn.com/602190090/files/doc_financials/2018/q3/q3-2018-earnings-webtables.pdf}
\BIBentrySTDinterwordspacing

\bibitem{hp-rev-q4-2018}
\BIBentryALTinterwordspacing
``Hp inc.. and subsidiaries consolidated condensed statements of earnings,''
  2018. [Online]. Available:
  \url{https://s2.q4cdn.com/602190090/files/doc_financials/2018/q4/q4-2018-earnings-webtables.pdf}
\BIBentrySTDinterwordspacing

\bibitem{hp-rev-q1-2019}
\BIBentryALTinterwordspacing
``Hp inc.. and subsidiaries consolidated condensed statements of earnings,''
  2019. [Online]. Available:
  \url{https://s2.q4cdn.com/602190090/files/doc_financials/2019/Q1/Q119-HP-Inc-Quarterly-Results.pdf}
\BIBentrySTDinterwordspacing

\bibitem{hp-rev-q2-2019}
\BIBentryALTinterwordspacing
``Hp inc.. and subsidiaries consolidated condensed statements of earnings,''
  2019. [Online]. Available:
  \url{https://s2.q4cdn.com/602190090/files/doc_financials/2019/Q2/Q219-HP-Inc.-Quarterly-Results-Web-Tables.pdf}
\BIBentrySTDinterwordspacing

\bibitem{hp-rev-q3-2019}
\BIBentryALTinterwordspacing
``Hp inc.. and subsidiaries consolidated condensed statements of earnings,''
  2019. [Online]. Available:
  \url{https://s2.q4cdn.com/602190090/files/doc_financials/2019/q3/q3-2019-earnings-webtables.pdf}
\BIBentrySTDinterwordspacing

\bibitem{hp-rev-q4-2019}
\BIBentryALTinterwordspacing
``Hp inc.. and subsidiaries consolidated condensed statements of earnings,''
  2019. [Online]. Available:
  \url{https://s2.q4cdn.com/602190090/files/doc_financials/2020/q1/Q120-HP-Inc.-Earnings-Press-Release.pdf}
\BIBentrySTDinterwordspacing

\bibitem{hp-rev-q1-2020}
\BIBentryALTinterwordspacing
``Hp inc.. and subsidiaries consolidated condensed statements of earnings,''
  2020. [Online]. Available:
  \url{https://s2.q4cdn.com/602190090/files/doc_financials/2020/q1/q1-2020-quarterly-results.pdf}
\BIBentrySTDinterwordspacing

\bibitem{hp-rev-q2-2020}
\BIBentryALTinterwordspacing
``Hp inc.. and subsidiaries consolidated condensed statements of earnings,''
  2020. [Online]. Available:
  \url{https://s2.q4cdn.com/602190090/files/doc_financials/2020/q2/q2-2020-quarterly-results.pdf}
\BIBentrySTDinterwordspacing

\bibitem{hp-rev-q3-2020}
\BIBentryALTinterwordspacing
``Hp inc.. and subsidiaries consolidated condensed statements of earnings,''
  2020. [Online]. Available:
  \url{https://s2.q4cdn.com/602190090/files/doc_financials/2020/q3/Q320-HP-Inc-Earnings-Press-Release.pdf}
\BIBentrySTDinterwordspacing

\bibitem{hp-rev-q4-2020}
\BIBentryALTinterwordspacing
``Hp inc.. and subsidiaries consolidated condensed statements of earnings,''
  2020. [Online]. Available:
  \url{https://s2.q4cdn.com/602190090/files/doc_events/2020/Q42020/Q4FY20-HP-Inc.-Press-Release-vFINAL.pdf}
\BIBentrySTDinterwordspacing

\bibitem{hp-rev-q1-2021}
\BIBentryALTinterwordspacing
``Hp inc.. and subsidiaries consolidated condensed statements of earnings,''
  2021. [Online]. Available:
  \url{https://s2.q4cdn.com/602190090/files/doc_events/2021/Q1/Q1FY21-Press-Release-FINAL.pdf}
\BIBentrySTDinterwordspacing

\bibitem{hp-rev-q2-2021}
\BIBentryALTinterwordspacing
``Hp inc.. and subsidiaries consolidated condensed statements of earnings,''
  2021. [Online]. Available:
  \url{https://s2.q4cdn.com/602190090/files/doc_events/2021/Q2/Q2FY21-HP-Inc.-Press-Release.pdf}
\BIBentrySTDinterwordspacing

\bibitem{hp-rev-q3-2021}
\BIBentryALTinterwordspacing
``Hp inc.. and subsidiaries consolidated condensed statements of earnings,''
  2021. [Online]. Available:
  \url{https://s2.q4cdn.com/602190090/files/doc_events/2021/Q3/Q3FY21-HP-Inc.-Press-Release.pdf}
\BIBentrySTDinterwordspacing

\bibitem{hp-rev-q4-2021}
\BIBentryALTinterwordspacing
``Hp inc.. and subsidiaries consolidated condensed statements of earnings,''
  2021. [Online]. Available:
  \url{https://s2.q4cdn.com/602190090/files/doc_financials/2021/q4/Q421-HP-Inc.-Press-Release.pdf}
\BIBentrySTDinterwordspacing

\bibitem{t4}
\BIBentryALTinterwordspacing
``Semiconductor factory fire in japan adds to global shortage,'' Jun 2021.
  [Online]. Available:
  \url{https://autovista24.autovistagroup.com/news/semiconductor-factory-fire-japan-adds-global-shortage/}
\BIBentrySTDinterwordspacing

\bibitem{auto}
\BIBentryALTinterwordspacing
``Semiconductor shortages causing automakers to cut production for early
  2021,'' Jan 2021. [Online]. Available:
  \url{https://info.fusionww.com/blog/semiconductor-shortages-causing-automakers-to-cut-production-for-early-2021}
\BIBentrySTDinterwordspacing

\bibitem{t6}
\BIBentryALTinterwordspacing
``Tsmc (tsm) - revenue,'' Oct 2021. [Online]. Available:
  \url{https://companiesmarketcap.com/tsmc/revenue/}
\BIBentrySTDinterwordspacing

\bibitem{amd_margin}
\BIBentryALTinterwordspacing
``Amd gross margin 2010-2022: Amd,'' Sep 2022. [Online]. Available:
  \url{https://web.archive.org/web/20221122070139/https://www.macrotrends.net/stocks/charts/AMD/amd/gross-margin}
\BIBentrySTDinterwordspacing

\bibitem{chips_cause_infl}
\BIBentryALTinterwordspacing
``Did the computer chip shortage affect inflation?'' May 2022. [Online].
  Available:
  \url{https://www.stlouisfed.org/on-the-economy/2022/may/did-computer-chip-shortage-affect-inflation}
\BIBentrySTDinterwordspacing

\bibitem{hp-rev-q1-2022}
\BIBentryALTinterwordspacing
``Hp inc.. and subsidiaries consolidated condensed statements of earnings,''
  2022. [Online]. Available:
  \url{https://s2.q4cdn.com/602190090/files/doc_financials/2022/q1/Q1FY22-HP-Inc.-Press-Release.pdf}
\BIBentrySTDinterwordspacing

\bibitem{hp-rev-q2-2022}
\BIBentryALTinterwordspacing
``Hp inc.. and subsidiaries consolidated condensed statements of earnings,''
  2022. [Online]. Available:
  \url{https://s2.q4cdn.com/602190090/files/doc_financials/2022/q2/Q2FY22-HP-Inc.-Press-Release.pdf}
\BIBentrySTDinterwordspacing

\bibitem{intc_margin}
\BIBentryALTinterwordspacing
``Intel profit margin 2010-2022: Intc,'' Sep 2022. [Online]. Available:
  \url{https://www.macrotrends.net/stocks/charts/INTC/intel/profit-margins}
\BIBentrySTDinterwordspacing

\bibitem{noauthor_results_2022}
\BIBentryALTinterwordspacing
``\BIBforeignlanguage{en}{Results from {Semiconductor} {Supply} {Chain}
  {Request} for {Information}},'' Jan. 2022. [Online]. Available:
  \url{https://www.commerce.gov/news/blog/2022/01/results-semiconductor-supply-chain-request-information}
\BIBentrySTDinterwordspacing

\bibitem{t8}
\BIBentryALTinterwordspacing
E.~Barrett, ``Semiconductors are a us-china trade war weapon. can tsmc serve
  both sides?'' Aug 2020. [Online]. Available:
  \url{https://fortune.com/2020/08/10/us-china-trade-war-semiconductors-chips-tsmc-chipmakers/}
\BIBentrySTDinterwordspacing

\bibitem{fab_takes_long_time}
\BIBentryALTinterwordspacing
H.~Bauer, O.~Burkacky, P.~Kenevan, S.~Lingemann, K.~Pototzky, and B.~Wiseman,
  ``Semiconductor design and manufacturing: Achieving leading-edge
  capabilities,'' Aug 2020. [Online]. Available:
  \url{https://www.mckinsey.com/industries/advanced-electronics/our-insights/semiconductor-design-and-manufacturing-achieving-leading-edge-capabilities}
\BIBentrySTDinterwordspacing

\bibitem{begen_supply_2016}
\BIBentryALTinterwordspacing
M.~A. Begen, H.~Pun, and X.~Yan, ``Supply and demand uncertainty reduction
  efforts and cost comparison,'' \emph{International Journal of Production
  Economics}, vol. 180, pp. 125--134, 2016. [Online]. Available:
  \url{https://www.sciencedirect.com/science/article/pii/S0925527316301591}
\BIBentrySTDinterwordspacing

\bibitem{t12}
\BIBentryALTinterwordspacing
R.~Beggin, ``As tensions simmer over taiwan, global chip supply hangs in the
  balance,'' Nov 2021. [Online]. Available:
  \url{https://www.proquest.com/wire-feeds/as-tensions-simmer-over-taiwan-global-chip-supply/docview/2591857351/se-2?accountid=14553}
\BIBentrySTDinterwordspacing

\bibitem{comp}
\BIBentryALTinterwordspacing
D.~Bonderud, ``How supply chain woes are affecting pc vendors and what’s next
  for the industry.'' [Online]. Available:
  \url{https://www.spiceworks.com/tech/hardware/articles/supply-chain-impacting-pc-vendors/}
\BIBentrySTDinterwordspacing

\bibitem{t3}
\BIBentryALTinterwordspacing
K.~Carlson, ``Nxp could lose \$100 million due to weather shutdown of austin
  plants,'' Mar 2021. [Online]. Available:
  \url{https://www.statesman.com/story/business/2021/03/12/nxp-could-lose-100-million-due-weather-shutdown-austin-plants/4664621001/}
\BIBentrySTDinterwordspacing

\bibitem{t2}
\BIBentryALTinterwordspacing
K.~Carlson, ``Shutdown of austin fab during freeze cost samsung at least \$268
  million,'' Apr 2021. [Online]. Available:
  \url{https://www.statesman.com/story/business/2021/04/30/austin-fab-shutdown-during-texas-freeze-cost-samsung-millions/4891405001/}
\BIBentrySTDinterwordspacing

\bibitem{mspr2}
C.~Condevaux-Lanloy and E.~Fragniere, ``Setstoch: A tool for multistage
  stochastic programming with recourse,'' in \emph{Logilab, University of
  Geneva}.\hskip 1em plus 0.5em minus 0.4em\relax Citeseer, 1998.

\bibitem{IDC_Semi_Rev_Forc_21}
\BIBentryALTinterwordspacing
S.~Endicott, ``Semiconductor revenue expected to reach \$522 billion this
  year.'' [Online]. Available:
  \url{https://www.windowscentral.com/semiconductor-revenue-expected-reach-522-year}
\BIBentrySTDinterwordspacing

\bibitem{binning}
\BIBentryALTinterwordspacing
N.~Evanson, ``Explainer: What is chip binning? hitting the silicon lottery
  jackpot.'' [Online]. Available:
  \url{https://www.techspot.com/article/2039-chip-binning/}
\BIBentrySTDinterwordspacing

\bibitem{clim_chan_drought_fire}
\BIBentryALTinterwordspacing
R.~Fowler, ``Drought and fire activity: What's climate change got to do with
  it?'' Aug 2016. [Online]. Available:
  \url{https://lamont.columbia.edu/news/drought-and-fire-activity-whats-climate-change-got-do-it}
\BIBentrySTDinterwordspacing

\bibitem{gilbert_2017}
\BIBentryALTinterwordspacing
J.~Gilbert, ``Partial equilibrium analysis part i a basic partial equilibrium
  ... - escap,'' May 2017. [Online]. Available:
  \url{https://unescap.org/sites/default/files/09_PE_I.pdf}
\BIBentrySTDinterwordspacing

\bibitem{harding_2018}
\BIBentryALTinterwordspacing
S.~Harding, ``What is binning? a basic definition,'' Oct 2018. [Online].
  Available:
  \url{https://www.tomshardware.com/reviews/glossary-binning-definition,5892.html}
\BIBentrySTDinterwordspacing

\bibitem{amdnext}
\BIBentryALTinterwordspacing
N.~Hemsoth, ``Amd on why chiplets—and why now.'' [Online]. Available:
  \url{https://www.nextplatform.com/2021/06/09/amd-on-why-chiplets-and-why-now/}
\BIBentrySTDinterwordspacing

\bibitem{t10}
Hutcheson, ``Remarks at sia event: Big opportunities, looming challenges: The
  state of the u.s. semiconductor industry.'' 2020.

\bibitem{amdpaper}
A.~Kannan, N.~E. Jerger, and G.~H. Loh, ``Enabling interposer-based
  disintegration of multi-core processors,'' in \emph{2015 48th Annual IEEE/ACM
  International Symposium on Microarchitecture (MICRO)}.\hskip 1em plus 0.5em
  minus 0.4em\relax IEEE, 2015, pp. 546--558.

\bibitem{lee_semiconductors_2022}
\BIBentryALTinterwordspacing
B.~Lee, ``\BIBforeignlanguage{en-US}{Semiconductors: {The} {Supply} and
  {Demand} {Fluctuations} {\textbar} {Suntsu} {News}},'' Jun. 2022. [Online].
  Available:
  \url{https://suntsu.com/2022/06/15/semiconductors-the-supply-and-demand-fluctuations/}
\BIBentrySTDinterwordspacing

\bibitem{Allign_SC_strat_Prod_Unc}
H.~L. Lee, ``Aligning supply chain strategies with product
  uncertainties{\textbar} {California Management Review. Spring2002, Vol. 44
  Issue 3, p105-119.}''

\bibitem{lingo}
I.~Lindo~Systems, ``Lingo 16.0—optimization modeling software for linear,
  nonlinear, and integer programming,'' 2017.

\bibitem{t11}
\BIBentryALTinterwordspacing
K.~Lyons, ``Us tightens trade restrictions on chinese chipmaker smic.''
  [Online]. Available:
  \url{https://www.theverge.com/2020/9/26/21457350/us-tightens-trade-restrictions-china-chipmaker-smic}
\BIBentrySTDinterwordspacing

\bibitem{semi_lead_time_long}
\BIBentryALTinterwordspacing
D.~Martin, ``Semiconductor average lead time breaks half-year barrier,'' Apr
  2022. [Online]. Available:
  \url{https://www.theregister.com/2022/04/14/semiconductor_lead_times/}
\BIBentrySTDinterwordspacing

\bibitem{amdepycandryzen}
S.~Naffziger, N.~Beck, T.~Burd, K.~Lepak, G.~Loh, M.~Subramony, and S.~White,
  ``Pioneering chiplet technology and design for the amd epyc™ and ryzen™
  processor families : Industrial product,'' 06 2021, pp. 57--70.

\bibitem{tvlsi}
S.~Pal, D.~Petrisko, R.~Kumar, and P.~Gupta, ``Design space exploration for
  chiplet-assembly-based processors,'' \emph{IEEE Transactions on Very Large
  Scale Integration (VLSI) Systems}, vol.~28, no.~4, pp. 1062--1073, 2020.

\bibitem{mspr}
G.~Pantuso and T.~K. Boomsma, ``On the number of stages in multistage
  stochastic programs,'' \emph{Annals of Operations Research}, vol. 292, no.~2,
  pp. 581--603, 2020.

\bibitem{glut_of_chips}
\BIBentryALTinterwordspacing
J.~L. Person, ``Computer chips face toilet paper hoarding moment as shortage
  turns to glut,'' Jul 2022. [Online]. Available:
  \url{https://www.reuters.com/technology/computer-chips-face-toilet-paper-hoarding-moment-shortage-turns-glut-2022-07-12/}
\BIBentrySTDinterwordspacing

\bibitem{t5}
\BIBentryALTinterwordspacing
A.~Shilov, ``Amd expected to become tsmc's second largest customer,'' Mar 2021.
  [Online]. Available:
  \url{https://www.tomshardware.com/news/amd-tsmc-second-largest-customer}
\BIBentrySTDinterwordspacing

\bibitem{clim_chan_winter_storm}
\BIBentryALTinterwordspacing
C.~Sweeny, ``Texas storm offers glimpse of how climate change threatens public
  health,'' Feb 2021. [Online]. Available:
  \url{https://www.hsph.harvard.edu/news/features/texas-storm-offers-glimpse-of-how-climate-change-threatens-public-health/}
\BIBentrySTDinterwordspacing

\bibitem{ITIF}
\BIBentryALTinterwordspacing
S.~E.~I. Technology and I.~F. (ITIF), ``An allied approach to semiconductor
  leadership,'' pp. 1--57. [Online]. Available:
  \url{https://itif.org/publications/2020/09/17/allied-approach-semiconductor-leadership}
\BIBentrySTDinterwordspacing

\bibitem{t1}
\BIBentryALTinterwordspacing
E.~Udin, ``The shutdown of tsmc's 14a fab will cause losses of over \$28
  million,'' Apr 2021. [Online]. Available:
  \url{https://www.gizchina.com/2021/04/17/the-shutdown-of-tsmcs-14a-fab-will-cause-losses-of-over-28-million/}
\BIBentrySTDinterwordspacing

\bibitem{venkat2014harnessing}
A.~Venkat and D.~M. Tullsen, ``Harnessing isa diversity: Design of a
  heterogeneous-isa chip multiprocessor,'' in \emph{2014 ACM/IEEE 41st
  International Symposium on Computer Architecture (ISCA)}.\hskip 1em plus
  0.5em minus 0.4em\relax IEEE, 2014, pp. 121--132.

\bibitem{t7}
\BIBentryALTinterwordspacing
M.~Wayland, ``How covid led to a \$60 billion global chip shortage for the auto
  industry,'' Feb 2021. [Online]. Available:
  \url{https://www.cnbc.com/2021/02/11/how-covid-led-to-a-60-billion-global-chip-shortage-for-automakers.html}
\BIBentrySTDinterwordspacing

\end{thebibliography}

\end{document}